\documentclass[letterpaper, twocolumn]{article}
\usepackage[a4paper,margin=0.85in]{geometry}

\usepackage{times}
\usepackage{helvet}
\usepackage{courier}

\usepackage[hyphens]{url}

\usepackage{amsmath}
\usepackage{graphicx}
\usepackage{multirow}
\usepackage{booktabs}
\usepackage{makecell}
\usepackage{subcaption}

\usepackage[affil-it]{authblk}
\usepackage[numbers]{natbib}
\usepackage{caption}

\title{Detecting Coordinated Activities Through Temporal, Multiplex, and Collaborative Analysis}

\author[1]{Letizia Iannucci}
\author[2,3]{Elisa Muratore}
\author[4]{Antonis Matakos}
\author[1]{Mikko Kivelä}
\affil[1]{Department of Computer Science, Aalto University}
\affil[2]{Department of Mathematics, Trento University}
\affil[3]{Bruno Kessler Foundation}
\affil[4]{Department of Computer and Systems Sciences, Stockholm University }

\date{}

\begin{document}

\maketitle

\begin{abstract}
    In the era of widespread online content consumption, effective detection of coordinated efforts is crucial for mitigating potential threats arising from information manipulation. Despite advances in isolating inauthentic and automated actors, the actions of individual accounts involved in influence campaigns may not stand out as anomalous if analyzed independently of the coordinated group. Given the collaborative nature of information operations, coordinated campaigns are better characterized by evidence of similar temporal behavioral patterns that extend beyond coincidental synchronicity across a group of accounts.
    We propose a framework to model complex coordination patterns across multiple online modalities. This framework utilizes multiplex networks to first decompose online activities into different interaction layers, and subsequently aggregate evidence of online coordination across the layers. In addition, we propose a time-aware collaboration model to capture patterns of online coordination for each modality.
    The proposed time-aware model builds upon the node-normalized collaboration model and accounts for repetitions of coordinated actions over different time intervals by employing an exponential decay temporal kernel. We validate our approach on multiple datasets featuring different coordinated activities. Our results demonstrate that a multiplex time-aware model excels in the identification of coordinating groups, outperforming previously proposed methods in coordinated activity detection.
\end{abstract}

\section{Introduction}

The widespread use of social media as a primary information source makes it vital to detect unreliable or malicious actors who exploit the features of those platforms to intentionally manipulate public opinion, intensify societal divisions, and influence policy decisions~\citep{howell2013digital}. Specifically, the ability to rapidly distribute content and facilitate real-time interactions creates opportunities for the inauthentic amplification of misleading or inflammatory information. For these reasons, early detection of influence operations and information campaigns becomes imperative to prevent the spread of propaganda and the deepening of societal divisions, and to maintain the integrity of the information environment~\citep{carley2020social}.

In response to the alarming observation of coordinated activities tampering with crucial events, such as political elections~\citep{giglietto2020takes, linhares2022uncovering, ng2022online}, researchers have devoted significant efforts to address this challenge by categorizing individual accounts as information polluters, bots, or other types of inauthentic social actors~\citep{ferrara2016rise, kumar2017army, elmas2021ephemeral}. Detecting individual malicious users can be challenging, especially within bot networks where the activity of each bot is strategically calibrated to evade detection. However, it is the coordinated nature of their activities, rather than the isolated actions, that ultimately generates a significant impact on the social media platform. Recent research underscores how information operations exhibit strong characteristics of collaborative work~\citep{starbird2019disinformation}, making coordinated group behavior a better descriptor than focusing solely on individual actions or isolated elements~\citep{keller2020political}.

This shift in perspective makes it possible to expose coordinated efforts by identifying an anomalous degree of consistency within a group of accounts~\citep{keller2020political}. When analyzing group dynamics, it becomes evident that malicious actors exploit numerous opportunities for coordination on online platforms. These opportunities manifest in various coordination strategies, such as utilizing the same set of hashtags, engaging in the same discussions, amplifying identical narratives, or repeatedly sharing content from a specific account or external source~\citep{ng2022online, pacheco2021uncovering}.

The complexity of online coordination is further heightened by the ability of groups of accounts to coordinate across multiple dimensions simultaneously~\citep{ng2022online, magelinski2022synchronized}. Coordination may occur in the temporal dimension, involving activities at similar times or with similar frequency~\citep{keller2020political}, and it may extend to the social dimension when coordinating actors interact with the same set of users~\citep{magelinski2022synchronized}. Coordination can operate on the semantic level through the use of identical words, hashtags, or similar text~\citep{elmas2021ephemeral}, while the referral dimension can be employed to redirect users to the same external content or sources~\citep{giglietto2020takes}. Finally, the manipulation of metadata can be leveraged to create fabricated personas designed to resemble regular users, making them well-suited for deployment in information campaigns~\citep{ferrara2016rise}. The potential for simultaneous coordination across multiple dimensions effectively turns the detection of online coordinated behavior into a multi-dimensional problem~\citep{ng2022online}. Therefore, reliable detection of coordinated behavior necessitates a comprehensive approach that considers the time patterns of collaborative effort across different potential strategies of coordination. These assumptions lead to the following research question: 

\textbf{RQ. }Does integrating the temporal aspect and multiple modalities of coordination improve the detection and characterization of collaborative efforts?

We address the multi-dimensional nature of coordination by combining different potential strategies of coordination in a multiplex representation of online latent collaboration. The multiplex representation is able to accumulate evidence of coordination from diverse coordination modalities, such as hashtag usage, user interactions, narrative amplification, and content sharing, and to describe latent collaboration on each of them. A multiplex model provides a rich description of coordinated activities, from which we can extract clusters of coordinating accounts that exhibit collaboration across different modalities.

As our second contribution, we propose an application of the node-normalized collaboration model introduced by~\citet{newman2001scientific} and we design it to model the significance of temporal patterns of online coordination. Coordinating accounts intentionally manipulate online spaces through common and repeated behavior~\citep{magelinski2022synchronized}. Thus, both the number of repetitions of common actions and their temporal differences represent fundamental indicators to distinguish coordination from random or independent user activity. In particular, similar actions occurring within a short time frame may signify malicious tactics, including spamming, participation in disinformation campaigns, or involvement in coordinated attacks~\citep{keller2020political}. Furthermore, repeated instances of the same activities, even if distributed over an extended period of time, suggest concerted efforts~\citep{weber2020s}. In both cases, larger weights assigned by the model capture the deliberate effort to synchronize or repeat activities beyond mere coincidence and highlight the collaborative nature of these actions.

To assess the efficacy of our proposed model, we begin by investigating its performance on a series of simple synthetic case studies. These controlled experiments allow us to validate the model's behavior against diverse patterns of coordination and to anticipate its strengths and limitations. We then benchmark the performance of our model against several state-of-the-art methods previously established for the analysis of coordinated campaigns, leveraging available labeled datasets~\citep{seckin2025labeled}. Our analysis demonstrates how our model contributes to advancing the identification of coordinated operations.

The remainder of the paper provides a review of related literature (Section~\ref{s:2}), introduces the model (Section~\ref{s:4}), and concludes with the presentation of experiments (Section~\ref{s:5}) and discussion of our findings (Section~\ref{s:6}).

\section{Related work}
\label{s:2}

\subsection{Coordination detection}
Coordinated inauthentic behavior refers to the collaborative actions of a group of social media accounts with the deliberate intent of deceiving other users~\citep{weber2021amplifying, pacheco2021uncovering}. This definition emphasizes the collective effort to purposely increase the visibility of specific content, regardless of the authenticity of the content being shared. Therefore, common methods for uncovering coordinated groups primarily rely on the study of synchronized or highly similar online actions that defy random occurrence~\citep{magelinski2022synchronized, ng2022online, linhares2022uncovering, pacheco2020unveiling}.

Such behavior can be studied through the analysis of latent coordination networks (LCNs)~\citep{weber2020s, weber2021amplifying}. An LCN is constructed by examining online activities within specific time-windows, and links between users are inferred from their shared or similar actions. Given that inauthentic coordination is often strategically orchestrated and designed to be covert, explicit relations or interactions, such as friendship or following relationships, might prove insufficient for detection~\citep{weber2020s, giglietto2020takes, varol2017online}. Thus, LCNs expand beyond direct interactions and capture hidden or implicit relations between users, allowing for the identification of digital behavior patterns that are similar beyond coincidence~\citep{weber2020s, pacheco2021uncovering, magelinski2022synchronized}. 

The implicit collaborative effort between users can be represented by a bipartite network where each account is connected to the digital actions they have performed. Such a bipartite network can then be reduced to an LCN by connecting users that have performed the same action. Thus, edges in the LCN can be weighted based on the strength of correlation between accounts, by considering factors such as repeated observations of the same action or behavior similarity. Metrics commonly employed to this scope include simple frequency, Jaccard similarity, cosine similarity, $\chi^2$, or TF-IDF~\citep{weber2020s, pacheco2021uncovering, magelinski2022synchronized, ng2022online, tardelli2024multifaceted}. 

Once the (weighted) user-to-user network is available, it becomes possible to extract significant patterns of implicit collaboration among the accounts. This is usually done by removing less significant coordination edges by applying backbone extraction methods before extracting tightly connected clusters~\citep{weber2020s, ng2022online} or focusing on the connected components~\citep{giglietto2020takes, pacheco2020unveiling}. Network analysis techniques such as community detection algorithms~\citep{weber2020s, pacheco2021uncovering, linhares2022uncovering}, or clustering~\citep{magelinski2022synchronized} are also frequently applied to identify cohesive groups within these networks.

While these methods might be effective, they typically focus on a single modality of coordination within a specific information operation event. Recognizing that users can coordinate across many strategies simultaneously, researchers have recently proposed multimodal coordination detection approaches. These methods generally fall into two categories: aggregating modalities into a single network, or retaining the distinct modalities through a multiplex representation~\citep{mannocci2025multimodal, graham2024coordination, magelinski2020detecting}.

The first approach involves combining evidence of coordination from different modalities into a single network, which allows for the application of well-established community detection algorithms and network centrality measures. This can be implemented by constructing a separate LCN for each modality of coordination, such as co-hashtags and co-mentions, and then extracting coordinated communities from the single weighted network obtained by a weighted sum of the edges found in each modality~\citep{weber2021amplifying}. Similarly, a metric to measure synchronization between any two users can be derived from the sum over different action types (e.g. URL, hashtag, mention)~\citep{ng2022combined}. Instead of summing, another variant of aggregation over modalities considers whether there is evidence of coordination in at least one modality~\citep{ng2023you, luceri2024unmasking, dey2024coordinated}. While this ensures that coordination in any single modality is captured, the model becomes vulnerable to tactics of dilution: malicious actors can perform a high number of irrelevant actions (i.e. coordinate at random with other users) to obscure their true coordinated group within the aggregated data.

To overcome this limitation, a second class of multimodal methods preserves the distinction between different types of coordination. This can be achieved using edge-colored networks, where each type of coordination (e.g., co-URLs, co-retweets, co-replies) is represented by a distinct edge color~\citep{graham2024coordination}. Alternatively, each atomic behavior, such as each individual URL or hashtag, can be encoded on a separate network layer~\citep{magelinski2020detecting, magelinski2022synchronized}. Indeed, retaining the multiplexity leads to greater robustness than clustering on aggregated networks, as multiplex clustering not only correctly identifies the most central nodes but also uncovers additional coordinated structures that aggregated networks would miss~\citep{magelinski2022synchronized, mannocci2025multimodal}.

However, it is worth noting that many current methods, both monomodal~\citep{weber2020s, ng2022online, pacheco2021uncovering} and multimodal~\citep{magelinski2020detecting, magelinski2022synchronized, weber2021amplifying, ng2022combined, ng2023you}, address the time dimension through fixed-size time windows, thereby limiting their adaptability to varied time strategies of coordination. For instance, several multimodal approaches, like the analysis  in~\citet{mannocci2025multimodal}, aggregate signals by simply averaging the weights of each modality over sliding time windows (e.g., 6 hours), without explicitly modeling the time component. Similarly, \citet{nwala2023language} encode user activity across multiple modalities (text, mentions, URLs, reshares, and others) into time-agnostic TF-IDF vectors derived from a formal language. While this representation can highlight similarities in behavioral patterns across modalities, it does not explicitly capture co‑occurrences and coordination in time. Also, an important limitation of fixed-window methods over aggregated modalities is that they inherently preclude the modeling of different time scales for different modalities, assuming a single, uniform time scale of coordination across all modalities. Methods that do model the temporal dimension, such as the multiplex approach by~\citet{tardelli2024temporal}, also still operate by specifying a fixed time window length and time window step. While ~\citep{tardelli2024temporal} focus on the temporal evolution of coordinated communities, they only consider co-retweet as modality of coordination. This approach not only overlooks other potential modalities, but also necessitates an empirical search over multiple time window lengths and steps to identify the appropriate time scale of coordination in the data. This empirical search is only possible when labeled data is available, which is not the case if we are interested in the early detection of new information operations. To expand our understanding of online coordination, we aim to analyze user behavior patterns across multiple modalities and to detect coordinated behavior across different potential time scales of coordination.

Beyond network-based approaches, machine learning (ML) and deep learning (DL) techniques represent another prominent category of approaches for the detection of coordinated behavior. Traditional ML approaches typically rely on a wide array of carefully engineered features, including behavioral patterns (e.g., posting frequency), linguistic cues (e.g., sentiment, specific keywords), metadata (e.g., account creation date, number of friends, number of followers), and graph-based features derived from interaction networks~\citep{ferrara2016rise, davis2016botornot, varol2017online}. However, these methods mainly focus on bot detection only, and they encounter significant challenges related to generalization, as the increasingly widespread use of AI and LLM blurs the distinction between automated and human activity~\citep{yang2019arming, nwala2023language}.

More recently, deep learning models, particularly Recurrent Neural Networks (RNNs) with attention layers for sequential behavior analysis, and Graph Neural Networks (GNNs) for analyzing complex network structures, have shown promising results~\citep{sharma2021identifying}. GNNs, in particular, are well-suited for modeling relationships between users and content, offering a powerful way to identify coordinated groups by learning directly from the network topology and node features~\citep{minici2025iohunter}. While powerful, these approaches often face challenges related to data imbalance, the need for extensive labeled datasets, the need for hyperparameter optimization, explainability of the models, and the dynamic, adversarial nature of inauthentic behavior which can lead to concept drift~\citep{luceri2024leveraging, minici2025iohunter}.

\subsection{Collaboration model}
Collaborative relationships between individuals have often been described through a collaboration network~\citep{ramasco2004self, melin1996studying, newman2001scientific, newman2001structure}. This mathematical model conceptualizes individuals as nodes within a graph, where edges signify collaborative connections, such as actors starring in the same movie, board members holding a position on the same corporate board, or researchers authoring the same publication.  For instance, in a co-authorship network, each node corresponds to an author, while an edge between two nodes signifies the existence of at least one publication co-authored by those individuals~\citep{melin1996studying, newman2001scientific, newman2001structure, newman2004coauthorship, moody2004structure, tahmooresnejad2018importance}.

Besides showing the existence of collaboration, some co-authorship networks include weighted edges to represent the strength of the collaboration between two individuals~\citep{melin1996studying, newman2001scientific, newman2001structure, newman2004coauthorship, ogasawara2025three}. While different researchers may adopt slightly different definitions for the weights, the collaboration model proposed by \citet{newman2001scientific} also takes care of normalization by defining the weight as:

\begin{equation}
    \label{eq:newmann}
    w_{uv} = \sum_k \frac{\delta^k_u \delta^k_v}{n_k - 1},
\end{equation}
where $\delta^k_u \delta^k_v$ = 1 if and only if both researchers have authored article $k$, which is co-authored by $n_k$ scientists in total. Hence, the strength of collaboration between any two authors is determined by the amount of their joint articles contributions, but the significance of the collaboration decreases with the number of co-authors on publication $k$.

Collaboration models have been used to map the structure of collaborating groups, for example to study the evolution of research communities, the emergence of interdisciplinary collaborations, and the identification of influential researchers and research groups~\citep{newman2004coauthorship, moody2004structure, tahmooresnejad2018importance, ogasawara2025three}. A collaboration model can also capture latent collaboration arising from shared activities within the context of online social media. In the digital adaptation, the shared activities can encompass various online interactions such as mentions, replies, retweets, or otherwise disseminating similar content. Therefore, by this model, social media users engaging in at least one similar or identical action will be marked as collaborating (i.e., will be linked by an edge).

\subsection{Multiplex networks}
Multiplex networks extend traditional graph models by allowing for multiple layers or modalities of interactions between nodes~\citep{lee2015towards}. A multiplex network can be reduced to an edge-colored network $G = \{ V, E, C\}$, where the edges $E \subseteq V \times V \times C$ support $|C|$ different types of relations between the nodes $V$~\citep{kivela2014multilayer}. This framework is particularly valuable in the analysis of online social media platforms, where users engage in diverse interactions like retweeting, tagging, befriending, following, or sharing the same content. 

Unlike aggregated networks, which collapse all interaction types into a single layer, a multiplex representation preserves the distinct nature and specific context of each type of relation. This is critical because different interaction modalities might carry varying semantic meanings, signal different types of social or behavioral connections, and importantly, operate across different time scales. For example, a reshare of a message is commonly understood as a rapid endorsement and amplification of content, while a mention or a reply signify direct conversational engagement, that can denote either agreement or disagreement~\citep{kwak2010twitter, honey2009beyond}.

\section{Model}
\label{s:4}

\begin{figure*}[!ht]
    \centering
    \includegraphics[width=0.8\textwidth]{figures/combined_schem_smaller.png}
    \caption{Our method for detecting coordinated behavior encodes user activities, with their timestamps, action types, and content, into a user-to-user multiplex network, where each layer models latent collaboration for a specific action type. Clusters of highly coordinating users are then extracted from the multiplex network (Panel \textbf{a}). The proposed time-aware model requires computing temporal differences ($\Delta t$) between users coactions. We consider the time differences between one user's action and the first subsequent occurrence of the same action by the other user (Panel \textbf{b}). The idea is that users act either as a response to external coordination or influenced by the activities from others: for example, they may use the hashtag they have just seen on their feed. The direction of influence is shown by the arrows: \textit{joe} might have used hashtag \textit{\#a} after seeing \textit{john}'s post at 5:55 or at 6:00, or \textit{john} might have decided to use hashtag \textit{\#a} at 6:15 after seeing \textit{joe}'s post at 6:02.}
    \label{fig:combined_schem}
\end{figure*}

\textbf{Notation.} Given an online social-media platform, we define the user base as a finite set
\(U = \{u_{1},\dots,u_{n}\}\) and let \(T\) denote the domain of timestamps of all user interactions. Interactions can be of various types, for example the creation of a profile, a post, or
the fact that one user followed another.
Let \(A = \{a_{1},\dots,a_{m}\}\) be the finite set of identifiable action
types on the platform.
For example, $ A = \bigl\{\text{\textit{tweet}},\;
              \text{\textit{retweet}},\;
              \text{\textit{hashtag}},\;
              \text{\textit{mention}},\;
              \text{\textit{url}},\;
              \dots\bigr\}.$

From each user interaction we derive a set of action-level data points
\[
  D = \{(u, t, a, k)\},
\]
where every tuple encodes the user \(u \in U\) performing the action,
the timestamp \(t \in T\), and the action type \(a \in A\). Finally, \(k \in K_a\) is the specific content (in the form of text string, RGB vector of the image, etc.) relevant to \(a\). As an example, a single post from user \textit{Alice} at time \(t\) might yield both a
\textit{hashtag} action with content \texttt{\#foo} and a
\textit{mention} action with content \texttt{@john}. Both data points are associated to the same user (\textit{Alice}, the author of the post) and have the same timestamp \(t\) referring to the publication of the post.

\textbf{Modeling assumptions.} We hypothesize that the temporal dimension of users activities is fundamental for identifying users engaging in coordination. While exact simultaneity may signal coordination, we assume that the significance of a co-occurring action between two users decreases as the time gap between them grows. However, as seen in previous research, a significant frequency of co-actions also represents evidence of coordination~\citep{keller2020political,weber2020s}. Thus, both temporal proximity and co-action frequency need to be considered in the modeling of coordination. 

Furthermore, we posit that separating modalities is essential for three reasons. First, it prevents the model from incorrectly equating different types of actions just because they occur at similar times. This preserves the unique semantic meaning of each action, which would otherwise be lost in an aggregated model. Second, this approach makes our model more resilient to tactics used by malicious actors. In fact, malicious actors could dilute the signal of their coordination by performing numerous irrelevant actions on random targets on one modality (for example, liking or following), while attempting to manipulate the social media on another modality (for example, by sharing the same sets of hashtags). Finally, not all actions hold equal weight in terms of their impact on the information space. For instance, sharing of the same content or direct engagement with certain users may exert a more significant influence than, for example, simple follows or likes, as they directly contribute to content amplification and targeted narrative shaping. For these reasons, we proceed to model different types of actions separately. 

\textbf{Time-aware collaboration model.} Given that all online activities are timestamped, the node-normalized collaboration model~\citep{newman2001scientific} can be extended to consider the time differences between occurrences of the same online action performed by two different users.  Let $T_{u}^k$ be the set of timestamps of user $u$ over content $k$. We define the multiset of time differences $\Delta T_{uv}^k$ between users $u$ and $v$ on content $k$ as the union of two multisets: the multiset $\Delta T_{u \to v}^k$ of time differences computed from user $u$'s perspective, and the multiset $\Delta T_{v \to u}^k$ of time differences computed from user $v$'s perspective:
\[
\begin{split}
    \Delta T_{u \to v}^k = \Big  \{ \min_{t' \in T_{v}^k, t' \geq t} (t' - t) \; \Big | \; t \in T_{u}^k, \;  \exists t' \in T_{v}^k : t' \geq t \Big \}.
\end{split}
\]
$\Delta T_{u \to v}^k$ can be understood as the set of time lags between user $u$, who performed the action first, and the closest subsequent matching action of the other user (see Figure~\ref{fig:combined_schem}b). We then take the union of the two sets:
\[
    \Delta T_{uv}^k = \Delta T_{u \to v}^k \cup \Delta T_{v \to u}^k.
\]
We adopt an exponential decay temporal kernel to weight the significance of the time differences, as online activities occurring closer in time are typically more indicative of coordination than those separated by longer durations. Mathematically, given an action type $a \in A$, we express the weight $w^a_{uv}$ of the edge representing coordination between users $u$ and $v$ as:

\begin{equation}
    w^a_{uv} = \sum_{k \in K_a} \sum_{\Delta t \in \Delta T_{uv}^k} e^{-\beta_a \Delta t} \frac{\delta^k_u \delta^k_v}{n_k - 1},
    \label{eq:weights}
\end{equation}
where the summation is both over different shared content ($k \in K_a$) and also over the time differences ($\Delta t \in \Delta T_{uv}^k$) between common and subsequent online action. Similarly to the collaboration model in Equation~\ref{eq:newmann}, $n_k$ represents the total number of users that have performed the action with content $k$, without considering repetitions. Thus, given one action type $a$, this formulation allows for considering various co-actions, as well as multiple instances of the same action over time, while discounting the contribution of actions that are generally popular among the users in the social media platform. As an example, we can model hashtag coordination by letting $k$ iterate over all the hashtags shared on the social media platform, while each $\Delta t$ represents one co-occurrence of one particular hashtag in the behavioral traces of users $u$ and $v$. Similarly, coordination on the referral dimension can be modeled by letting $k$ iterate over all the direct mentions or @-tags, while $\Delta t$ represents one co-occurrence of one particular @-tag in the behavioral traces of users $u$ and $v$.

For each action type $a$, the weight assigned to each instance of collaboration is determined by assuming an exponential decay with parameter $\beta_a \geq 0$, where the choice of a higher decay rate reflects the assumption that evidence of collaboration can be seen from co-actions happening over smaller time intervals. Conversely, a lower value of $\beta_a$ implies the assumption that coincidental activities occurring further apart in time still constitute valuable evidence of collaboration. A minimal example illustrating the effect of $\beta_a$ on the collaboration network is shown in Appendix~\ref{app:choice-of-beta} (Figure~\ref{fig:time-aware-model}).

Similarly to how the size of the time window is chosen in~\citet{ng2022online}, we choose the temporal decay parameter $\beta_a$ for each modality by maximizing the modularity of the resulting layer-specific graphs (see Appendix~\ref{app:choice-of-beta}). The idea is that the $\beta_a$ yielding the highest modularity best highlights the latent community structures that arise from temporal coordination.

From a computational perspective, the weight computation in our time‑aware collaboration model is efficient despite the added temporal dimension. Given the exponential decay kernel parameterized by $\beta_a$, the fact that contributions from co-actions beyond a certain $\Delta t$ become negligible can be exploited to keep the computational complexity of each layer below quadratic in the number of co-actions. The exponential decay in Equation~\ref{eq:weights} allows for an optimized implementation that ignores co-actions beyond a maximum interval $\Delta t_{\text{max}} = \frac{- \ln{\epsilon}}{\beta_a}$, where $\epsilon$ is a chosen tolerance level. Under this approximation, any omitted co-action pair contributes at most $\frac{\epsilon}{n_k - 1}$ to the total weight, and the resulting approximation is strictly bounded by the sum of ignored tails. Since $\frac{1}{n_k - 1} \leq 1$ for all co-actions, the total error is at most the chosen tolerance level $\epsilon$ multiplied by the count of omitted pairs. This approximation helps most when $\beta_a$ is large relative to inter-coaction times. While in theory very dense and bursty co‑actions could push the complexity again toward quadratic, in our experiments this was not an issue.

\textbf{Multiplex network.} Instead of aggregating the different modalities of coordination in the same model, we represent the multidimensionality of collaboration with a multiplex network.  As each layer $c \in C$ in a multiplex network encodes a specific type of interaction, we can construct one layer for each action type that we wish to include in our analysis ($C = A$). By doing so, the multiplex representation enables us to assign different $\beta_a$ values to each coordination modality, as the significance of the same $\Delta t$ between co-activities may vary depending on the specific activity. Finally, we can utilize multiplex community detection algorithms to uncover cluster structures that span a combination of layers and that may not be apparent when analyzing individual layers in isolation.

\textbf{Community detection.} Community detection applied to latent collaboration networks represents a powerful tool for identifying users that are likely to be coordinating their actions~\citep{ng2022online}. Similar to monoplex networks, community detection in multilayer networks aims at identifying clusters of nodes that are strongly connected across multiple dimensions~\citep{mucha2010community}. In pursuit of this objective, techniques based on network modularity and spectral clustering have been adapted for multiplex and multilayer networks~\citep{mucha2010community, pamfil2019relating, deford2019spectral}. These adaptations have given rise to methods such as the generalized Louvain algorithm, Leiden algorithm, tensor decomposition, and generalized random walk. These multiplex community detection methods facilitate the detection of communities that span across multiple layers, eliminating the necessity of compressing the multiplex network into a monoplex representation. In fact, simplifying the network structure by collapsing multiple layers into a single dimension might lead to information loss and potential inaccuracies regarding the significance of users interactions, especially when the edge weights are not comparable~\citep{weber2020s}. We utilize Leiden algorithm to extract clusters of coordinating accounts from the multiplex representation of the online social media activities. As Leiden algorithm is based on multislice modularity~\citep{mucha2010community}, it allows us to identify groups of accounts that exhibit strong internal connections, suggesting coordinated behavior, while having weaker connections with accounts outside the group~\citep{mucha2010community}. Since the edges derived with our time-aware model are weighted, we extract clusters of coordinating accounts with Leiden algorithm by optimizing the multislice weighted modularity~\citep{mucha2010community}. Figure~\ref{fig:combined_schem} summarizes our setup.

\section{Experiments}
\label{s:5}

\subsection{Metrics for model evaluation}

We evaluate the ability of the model to detect the predefined coordination patterns. We quantify this by interpreting each community as one potentially coordinated group, and comparing its members to the ground truth. We score each community by calculating the F1 score. This metric penalizes cases where inauthentic accounts are assigned to different detected communities, and therefore favors models that not only find some evidence of coordination but are also able to identify and group the coordinated group as a single cluster. This reflects how useful the model is for real-world applications: fragmented detection requires more manual effort to piece together the full picture of a campaign, or worse, could lead to overlooking parts of it. We define F1* as the score for the community having the highest F1 score, and we also consider precision and recall for the community having the highest F1 score, to assess how well our model identifies coordinated activities.

In addition to the performance of the best cluster, we also evaluate the results on two global metrics: homogeneity, and a variation of weighted precision. The homogeneity score of the full cluster labeling penalizes a model that outputs clusters with mixed memberships. A high homogeneity score indicates that each detected cluster is composed almost entirely of accounts from a single ground truth campaign, which is crucial for analysts needing to quickly investigate the clusters for evidence of information operations.

We then propose a custom weighted precision score (WP) that implements a weighted average of the positive rates for all non-singleton clusters:

\begin{equation}
    WP = \frac{\sum_k n_k p_k^2}{\sum_k n_k p_k},
\end{equation}
\label{eq:4}
where $n_k$ is the size of the $k$-th cluster and $p_k$ is the true positive rate of the $k$-th cluster if $n_k>1$ and $0$ otherwise. This metric is designed to achieve three goals. First, it assigns $0$ precision to singletons, ensuring that an over-fragmented clustering that fails to capture coordination as a group phenomenon receives a poor score. Second, by weighting larger clusters more heavily, it rewards clusters that are both large and highly precise, reflecting the strategic value of correctly detecting coordination threats in their entirety. Finally, unlike a standard completeness score, our measure does not penalize methods that partition authentic users into clusters, focusing its evaluation solely on the ability of the model to identify inauthentic groups.

Weighted precision also serves as a direct measure of the practical utility of the model in an analysis pipeline. It quantifies how effective the model is at guiding an analyst toward genuinely coordinated clusters. The term $n_k p_k$ approximates the number of true positives found within a cluster, and $p_k$ reflects the likelihood that an analyst would flag the entire $k$ cluster as suspicious after finding evidence of coordination from random sampling. The sum over over $n_k p_k^2$ gives an estimate of the total number of coordinated accounts an analyst can expect to find.

\subsection{Simulations}

\begin{figure*}[!ht]
  \centering
  \includegraphics[width=1\textwidth]{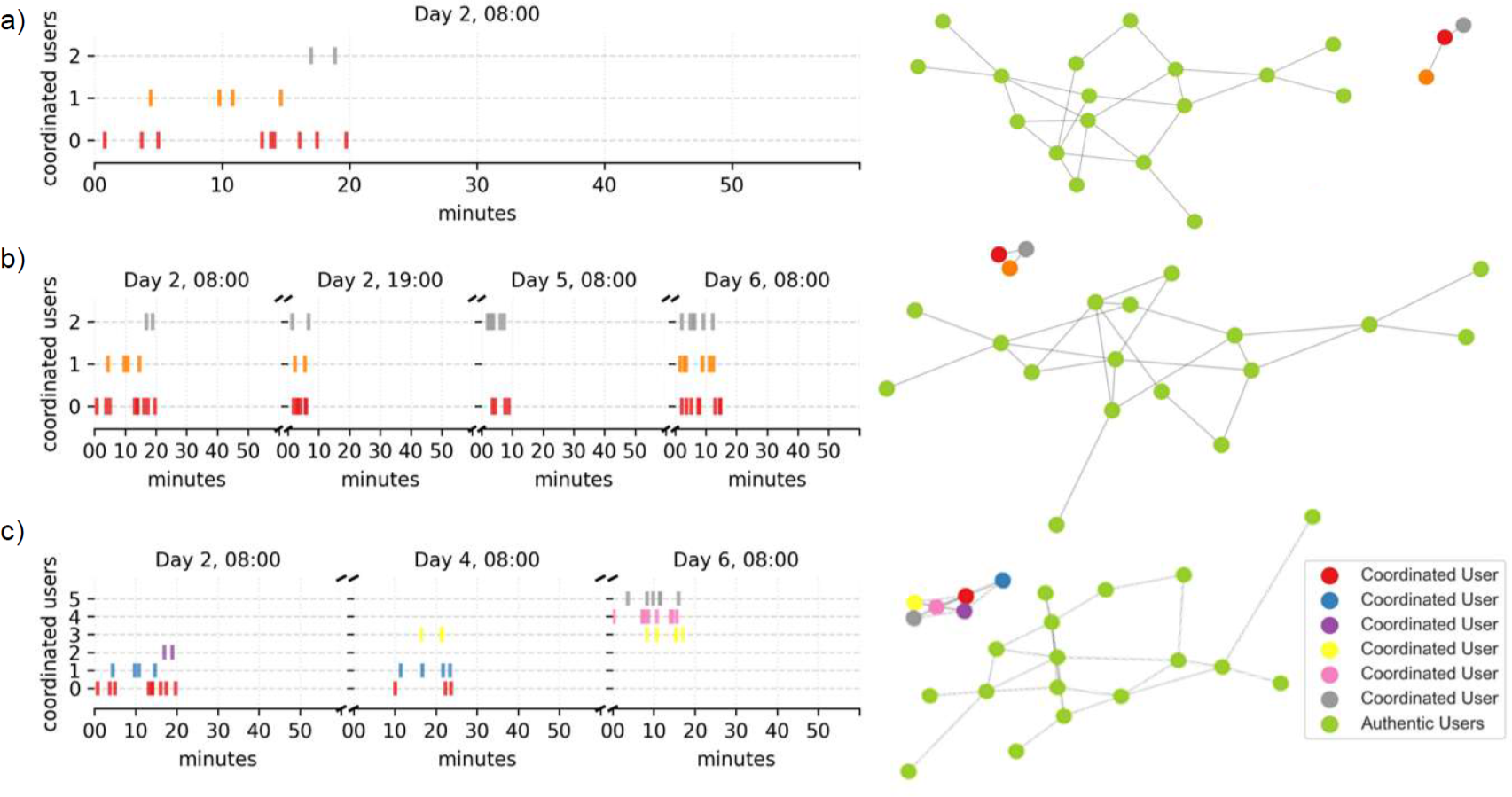}
  \caption{Time patterns of the simulations used to validate the model, and corresponding network layers for each of the simulated examples. Simulation 1 in \textbf{a)} creates almost synchronous bursts of activities; Simulation 2 in \textbf{b)} alternates bursts to periods without activities; Simulation 3 in \textbf{c)} alternates which inauthentic users are active at different times. Bursts of activities from inauthentic accounts are shown as vertical markers in the time event graphs. The corresponding network layers show the inauthentic accounts appearing as clear clusters after selecting $\beta_a$ via modularity maximization.}
  \label{fig:simulations}
\end{figure*}

To answer our research question, we first need to assess whether our model operates consistently with its design, and how well it is able to highlight coordination patterns. To this end, we employ simulations that replicate coordination strategies observed in real-world scenarios. By analyzing the performance of the model on these examples, we validate that it correctly captures coordination signals and that it can applied on real-world datasets. 

Each simulation mimicks a specific time pattern of coordination that can be potentially employed by coordinated groups on online platforms. In Simulation 1, coordinating accounts engage on a burst of intense activity characterized by closely timestamped actions. In Simulation 2, we simulate periods of intense activity alternating with periods of complete inactivity. In Simulation 3, we model the alternating behavior of active and silent accounts: the simulated coordinating actors exhibit activity over distinct time frames, with a fraction of the accounts remaining silent within each time frame. 

Figure~\ref{fig:simulations} illustrates, for each simulation, a subset of the time event graphs and monoplex networks constructed by maximizing modularity. Details on the simulations are reported in Appendix~\ref{app:sim-results}.

Our model achieves a perfect score of 1.00 across F1*, homogeneity, and weighted precision, with the exception of F1* for Simulation 3 in the monoplex version (see Appendix~\ref{app:sim-results}). This isolated shortcoming is due to the alternating users being split into two distinct clusters. However, this issue is resolved when considering more than one layer. Also, our model always outperforms a random labeler on the same number of clusters (see Appendix~\ref{app:sim-results}). This provides evidence that our time-aware multiplex model is robust to complex temporal patterns and effectively identifies coordination across different modalities.

\begin{figure*}[!ht]
    \centering 
    \includegraphics[width=1\textwidth]{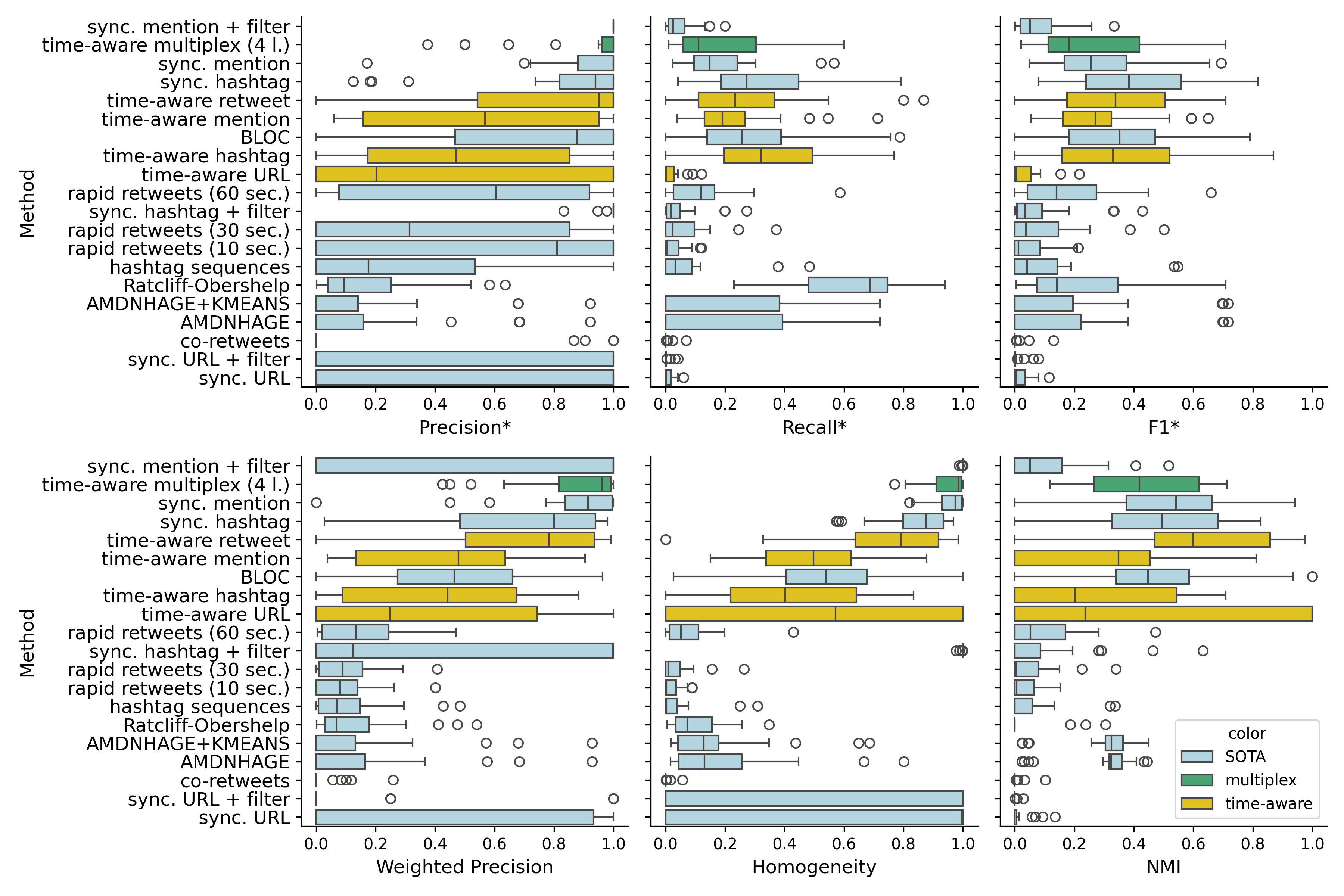}
    \caption{Results across the 26 labeled datasets provided by~\citet{seckin2025labeled} show that, in general, our time-aware model tends to achieve consistently better performance than many other approaches already on one layer when evaluated on the F1 score of the best cluster (F1*). The multiplex model not only achieves high precision* and homogeneity, but also the best overall weighted precision.}
    \label{fig:results-comparison}
\end{figure*}

\begin{figure}[!ht]
    \centering 
    \includegraphics[width=1\columnwidth]{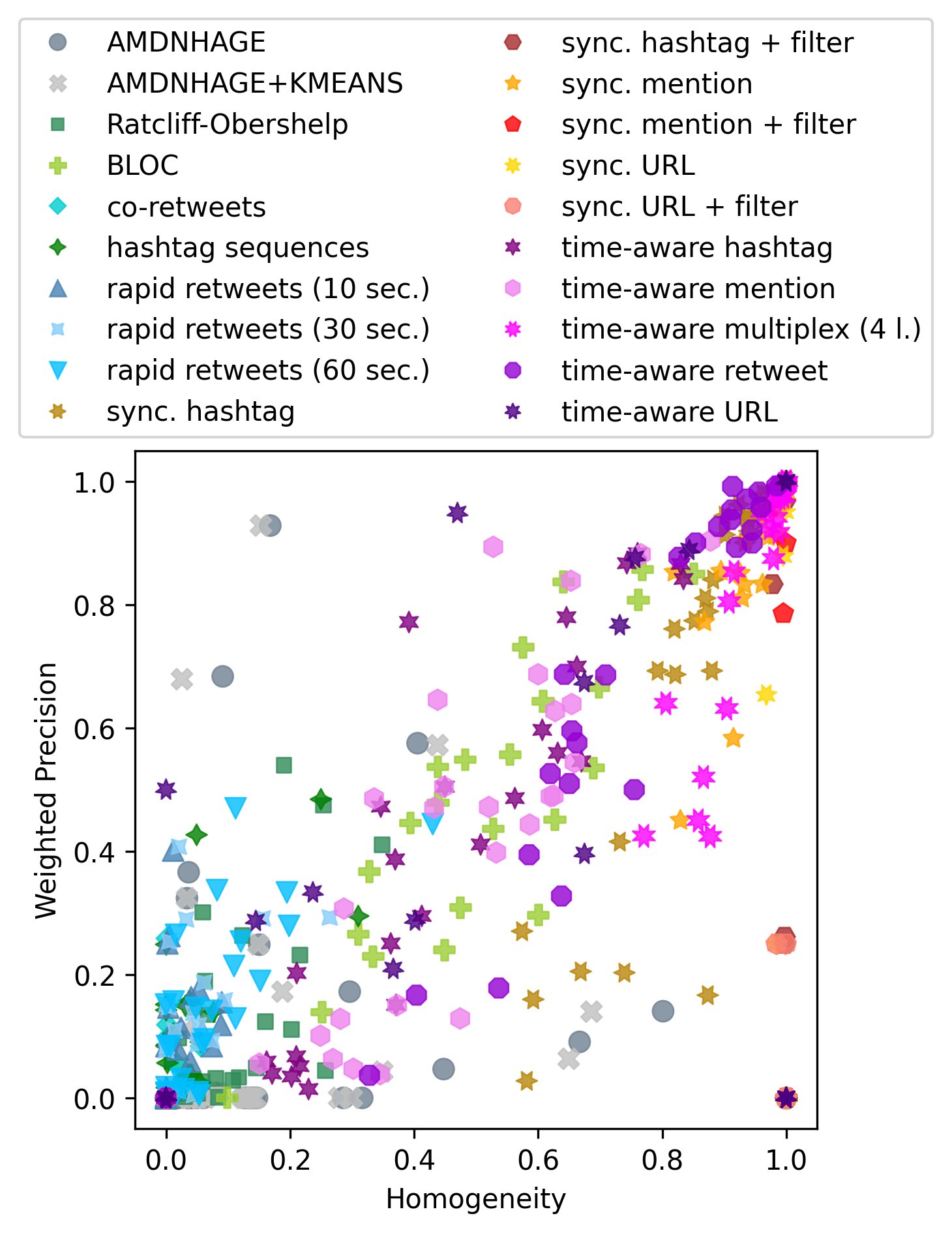}
    \caption{Weighted Precision and Homogeneity across the 26 labeled datasets provided by~\citet{seckin2025labeled}, showing that the time-aware multiplex method achieves both high weighted precision and high homogeneity.}
    \label{fig:results-scatter}
\end{figure}

\subsection{Benchmarking on labeled datasets}

Next, we evaluate the performance of our model by benchmarking it against other coordination detection methods on datasets with known ground truth~\citep{seckin2025labeled}. This step allows us to assess how well our approach identifies coordination patterns compared to existing techniques. We consider detection methods based on co-occurence of hashtag sequences~\citep{pacheco2021uncovering}, cardinality of co-retweets~\citep{ , linhares2022uncovering}, Ratclif-Obershelp similarity of text~\citep{pacheco2020unveiling, vishnuprasad2024tracking}, cardinality of co-actions across sliding time windows~\citep{magelinski2022synchronized, ng2022online}, and AMDN-HAGE and AMDN-HAGE+KMEANS models~\citep{sharma2021identifying}. We further include BLOC~\citep{nwala2023language}, in its unsupervised implementation, as a multimodal approach. This selection provides broad coverage of unsupervised methods, encompassing those that explicitly model user collaboration (from simple cardinality to similarity) as well as machine-learning based approaches (such as AMDN-HAGE).

We analyze precision, recall, and F1 scores for the best cluster, as well as homogeneity, Normalized Mutual Information (NMI), and our custom weighted precision, to ensure a fair comparison across different methods (see Appendix~\ref{app:full-results}). Figure~\ref{fig:results-comparison} shows the distribution of all scores for each method. Our results show considerable variability in the performance of methods accounting for only a single coordination modality. Their effectiveness heavily depends on whether the specific modality they consider is actively used in the coordinated behavior. A lack of activity in that particular modality often results in insufficient evidence for detection, which yields low precision and recall scores. For example, methods relying on cardinality of co-retweets failed to detect coordination in Iran campaigns. The inherent limitations when relying on a single modality also extend to our model when restricted to a single layer. However, a key advantage of our individual layers is their ability to detect communities exhibiting coordination across different temporal scales, which offers better robustness compared to methods that rely on pre-filtering the time window under analysis. This is evident in the good individual performance of our retweet, hashtag, and mention layers for campaigns that are based on text, such as Armenia, Iran, or Russia. Also, our retweet layer generally achieves better scores on all metrics compared to methods relying on rapid retweets within a fixed-sized time window.

As for the multiplex results, our method achieves high precision and homogeneity (Figure~\ref{fig:results-scatter}). While other approaches, such as the ones in~\citet{magelinski2022synchronized} and~\citet{ng2022online} also achieve high precision and homogeneity, they do so by finding trivial communities (singletons or very few users). Our multiplex model, on the other hand, also excels when scored with weighted precision (Equation~\ref{eq:4}), which discards trivial clusters. Moreover, the F1* obtained from our multiplex method is only surpassed by communities found on monoplex layers. This can be explained by the fact that communities that exist within a single layer could be further fragmented when that layer is combined with others in a multiplex clustering. This hypothesis is supported by the observation of high precision but small recall for the best cluster obtained by multiplex clustering on our time-aware layers. 

We also report the Normalized Mutual Information (NMI) to compare clustering results to the binary ground truth. For this evaluation, we first binarize the clustering output. We label an entire cluster as positive if the majority of its members originally belonged to the information operation. The rationale for this approach is that an analyst tasked with manual inspection would likely label an entire cluster as suspicious after repeatedly finding evidence of coordination when sampling the users' online activities. In essence, this  measures how well the output of a method (i.e. the clusters or the split between suspicious/non suspicious clusters) aligns with the real-world, pragmatic goal of helping an analyst identify suspicious groups. The NMI score, in this context, becomes a measure of the usefulness of the model for the specific analytical task.

Figure~\ref{fig:results-comparison} shows that our multiplex method achieves among the highest NMI scores. While the three highest median NMI scores are achieved by our monoplex retweet layer and by the approaches from~\citet{magelinski2022synchronized} on mentions and on hashtags, Figure~\ref{fig:results-comparison} reveals again the key limitation of approaches that rely on a single modality: their NMI scores can be highly dependent on the chosen modality and can drop to nearly zero when applied to an action type with a weak signal, such as URL sharing.

Our method, on the other hand, demonstrates superior robustness in two ways. First, the median NMI score of our monoplex URL layer is considerably higher than that of the URL approach from~\citet{magelinski2022synchronized}. This shows that our time-aware collaboration model is better at capturing even small signals of coordination. Second, our multiplex method achieves NMI scores comparable to the best monoplex approaches, and without requiring prior knowledge of which specific modality holds the signal of coordination. This makes our approach more suitable for real-world applications where the nature of the information operation is not known in advance.

Table~\ref{tab:results} compares the average ranks of coordination detection methods based on weighted precision and across all metrics. The time-aware multiplex model with 4 layers emerges as the most reliable approach under weighted precision and remains the second-best performer overall, underscoring the value of integrating temporal information across modalities. The individual monoplex layers also rank within the top half of methods, confirming that temporal signals are informative even when considered in isolation.

\setlength{\tabcolsep}{3pt}
\begin{table}[h]
\centering
\begin{tabular}{lcc}
Method & \makecell{Avg. Rank \\ (Weighted Prec.)} & \makecell{Avg. Rank \\ (All Metrics)}\\
\midrule
time-aware multiplex (4 l.) & \textbf{3.12} & \underline{5.10} \\
sync. mention & \underline{3.38} & \textbf{4.74}\\
sync. mention + filter & 5.27 & 7.05 \\
sync. hashtag & 5.58 & 5.18 \\
time-aware retweet & 5.85 & 5.67 \\
time-aware mention & 8.00 & 7.84 \\
BLOC & 8.23 & 7.06 \\
time-aware hashtag & 8.81 & 7.99 \\
sync. hashtag + filter & 8.88 & 8.10 \\
time-aware URL & 9.54 & 10.37 \\
sync. URL & 9.96 & 10.74 \\
rapid retweets (60 sec.) & 11.08 & 10.56 \\
Ratcliff-Obershelp & 12.19 & 10.53 \\
rapid retweets (30 sec.) & 12.77 & 12.58 \\
AMDNHAGE & 12.96 & 11.78 \\
sync. URL + filter & 13.04 & 12.81 \\
AMDNHAGE+KMEANS & 13.23 & 11.92 \\
hashtag sequences & 13.35 & 12.94 \\
rapid retweets (10 sec.) & 13.65 & 12.87\\
co-retweets & 15.35 & 15.04 \\
\bottomrule
\end{tabular}
\caption{Performance comparison of coordination detection methods by the average rank across the 26 labeled datasets provided by~\citet{seckin2025labeled}. An average rank of 1 would mean that the method is the top performer across all 26 datasets. Our time-aware multiplex method is the \textbf{best one}, and our individual time-aware layers outperform most of other methods. When considering all the 6 metrics in Figure~\ref{fig:results-comparison}, our multiplex method is the \underline{second best}.}
\label{tab:results}
\end{table}

\section{Discussion}
\label{s:6}

As ~\citet{starbird2019disinformation} argue, information operations are best understood as collaborative work, where the behaviors of malicious actors and the ones of organic crowd might intersect. Signals of coordination can therefore arise both from deliberately coordinated actors (e.g., troll farms, paid actors) and from unwitting participants who amplify narratives. Platform affordances also shape how coordination manifests, providing distinct modalities through which coordination signals emerge. Signals from different modalities, when integrated, provide a more robust basis to separate deliberate and covert collaboration from organic dynamics.

Building on this perspective, our work contributes to three aspects that are crucial for reliable detection of modern coordinated campaigns: first, understanding how to model time patterns of collaboration; second, addressing the various modalities of collaboration and their role in the overall coordinated effort; and third, performing extensive benchmarking of methods for coordination detection.

As demonstrated by F1* and weighted precision scores, our proposed time-aware collaboration model, combined with a multiplex representation of social media activities, isolates coordinating groups more efficiently than previously suggested methods. This enhanced effectiveness is largely attributable to our model's inherent ability to incorporate and analyze different layers, or modalities, of coordination. Unlike detection methods focused on a single type of interaction, our approach is significantly less susceptible to information campaigns that strategically operate exclusively within one modality. Our benchmarking results have shown that coordinated campaigns primarily leveraging hashtag sharing were not easily detected by methods solely reliant on retweet or URL co-sharing. However, for the crucial task of early detection of information operations, anticipating the precise modality inauthentic actors will exploit is often impossible. A key strength of our model is that it does not require analysts to have prior knowledge of either the modality or the time scale at which coordination occurs. Many of the benchmarked approaches rely on prior assumptions, such as deciding whether retweets, hashtags, or URL sharing are the most informative signals, or fixing a temporal resolution (e.g., 5‑minutes sliding time windows, or retweets within 30 seconds). Our time‑aware multiplex framework avoids any heavy assumption by integrating signals across modalities and dynamically adapting to temporal patterns. This flexibility makes the approach more robust to the evolving strategies employed in real information operations.

An important consideration, however, is the availability of data. Access to comprehensive, high‑quality datasets remains a crucial factor for the development and benchmarking of coordination detection frameworks, yet such datasets are often focusing only on one known type of information operation (e.g. rapid retweets, or co-hashtag) or they over-represent one particular social media platform. These constraints have direct methodological implications. When certain modalities are underrepresented or inaccessible, detection methods and their evaluation become biased toward the signals that remain observable. For instance, while retweets have been extensively leveraged as indicators of coordination, other modalities such as URL sharing or mentions are far less frequently represented in the literature. This imbalance not only undermines the evaluation of robustness but might also skews research efforts toward methods optimized for a narrow set of modalities. As a result, detection methods risk lacking generalization and applicability in the real world.

Nevertheless, the method we propose is sufficiently general to be adapted across platforms and over different modalities. As long as a platform provides the basic affordances of time‑stamped interactions (in the form of text posts, reposts, reshares, replies, comments, quotes, likes, or other signals), our time‑aware multiplex framework can be applied without any further assumption than the fact that, in order to manipulate the information landscape, malicious actors need to coordinate their narrative and actions in time. Such coordination inevitably leaves detectable traces across one or more modalities, whether in the text or through other engagement with users and sources. By leveraging these temporal multimodal signals, our framework remains broadly applicable and not tied to the particular characteristics of a single information operation or of a single platform.

\subsection{Limitations}
Even though our multiplex time-aware model demonstrates strong performance in detecting coordinated groups, several limitations remain. First, the clusters we obtain come without any characterization of their content. Users on online social platforms might be coordinating for social good or coordination might arise from large-scale events. Our framework will extract clusters of coordinating accounts, but the intervention of a human analyst is needed in order to distinguish potential malicious coordination from grassroot.

Second, our reliance on modularity-based community detection brings along well-known shortcomings~\citep{fortunato2007resolution, lancichinetti2011limits}. Modularity optimization in multiplex networks requires the choice of the resolution parameter and inter-layer coupling strength, both of which can substantially affect the resulting partitions. In sparse networks, modularity is prone to overfitting, amplifying weak signals and potentially producing spurious communities~\citep{fortunato2007resolution}. Moreover, modularity-based community detection methods always result in some partitioning of the nodes in the network, regardless of whether each community exhibits meaningful coordination. This means that also genuine accounts will always be assigned to some community, even if the level of coordination is quite low and arises by chance. Future work should explore approaches to characterize the coordinated groups based on statistical evidence of coordination that goes beyond random chance, and account for the possibility that accounts not participating in malicious coordination may not belong to any cluster or community.

While the model's ability to dynamically adapt to various temporal scales is a significant strength, its current reliance on the optimal choice of the $\beta_a$ parameter for the temporal decay represents an area for refinement. Although we have proposed to choose $\beta_a$ by maximizing the modularity of each layer, further improvements could be realized by learning $\beta_a$ from the underlying data. This could involve machine learning approaches to determine the optimal $\beta_a$ for each layer. Furthermore, future research could extend our proposed model by considering a non-constant $\beta_a$ parameter, allowing it to dynamically adapt to the temporal characteristics of the data rather than assuming a fixed exponential decay. For example, $\beta_a$ could be replaced by an instantaneous rate function, such as the ones derived from time-varying stochastic processes. These future directions will refine our current approach to modeling temporal patterns of coordination, which already excels at integrating signals across diverse modalities and adapting to various temporal scales.

\begin{small}
\bibliographystyle{unsrtnat}
\bibliography{main}
 \end{small}

\paragraph{Funding Statement.}
This research was supported by grants from the Research Council of Finland (357743).

\appendix

\section{Choice of $\beta_a$}
\label{app:choice-of-beta}

Our primary method for selecting the optimal $\beta_a$ for each layer relies on identifying the value that produces a network with the most well-defined community structure. We operationalize this by searching for the $\beta_a$ that maximizes the modularity score of the resulting network partition. Modularity (Q) is a quality function that measures the density of edges within communities as compared to what would be expected in a random network. A higher modularity score indicates a more significant and less coincidental community structure.

The procedure is as follows. For each layer of the multiplex network, we iterate through a range of $\beta_a$ values, typically from 0 to 10 with a step of 0.01. For each value of $\beta_a$, we compute the weights of the edges and construct the corresponding weighted network. We then apply the Leiden community detection algorithm to this network to find the optimal community partition, and calculate the modularity (Q) of this partition for the generated network.

This process yields a mapping from each potential $\beta_a$ to the modularity score. We then select the $\beta_a$ value that corresponds to the global maximum of this function. By choosing the $\beta_a$ that maximizes modularity, we are essentially allowing the data to guide us towards the correct time scale of the analysis, where the underlying group patterns of one layer, which may be partially hidden by noise, become maximally evident.

\begin{figure*}[!t]
  \centering
  \includegraphics[width=\textwidth]{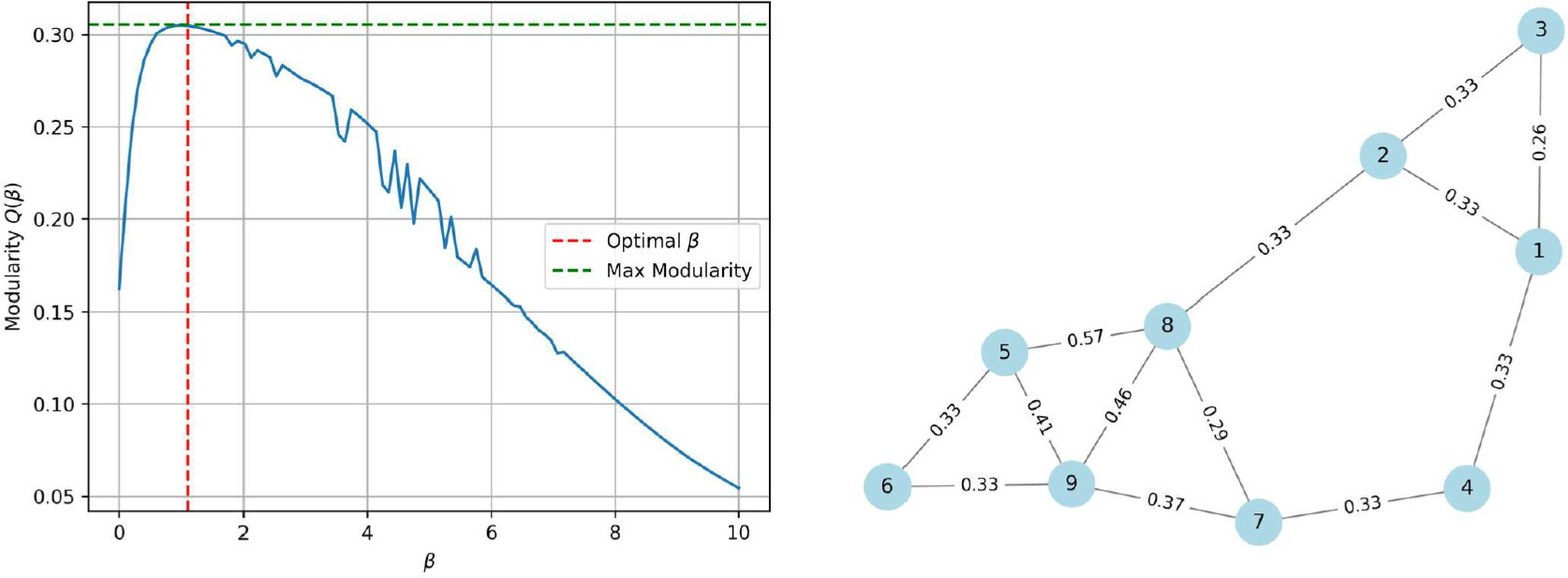}
  \caption{Modularity $Q$ as a function of $\beta$, and the network resulting from $\beta$ that maximizes modularity.}
  \label{fig:modularity-maximization}
\end{figure*}

As illustrated in Figure~\ref{fig:modularity-maximization}, which plots modularity against $\beta$ for an example monoplex network, the modularity score often exhibits a clear peak. This peak represents the optimal time resolution where the network better reveals its community structure.

The example in Figure~\ref{fig:modularity-maximization} utilizes a graph having the following edges, where the weight is parametrized by an exponential function with a parameter $\beta$:
\begin{center}
\begin{tabular}{@{}ccc@{}}
Node 1 & Node 2 & Edge Weight \\
1 & 2 & $e^{-1.0\beta}$ \\
1 & 3 & $e^{-1.2\beta}$ \\
1 & 4 & $e^{-1.0\beta}$ \\
1 & 7 & $e^{-7.0\beta}$ \\
1 & 9 & $e^{-5.0\beta}$ \\
2 & 3 & $e^{-1.0\beta}$ \\
2 & 7 & $e^{-5.0\beta}$ \\
2 & 8 & $e^{-1.0\beta}$ \\
4 & 5 & $e^{-7.5\beta}$ \\
4 & 6 & $e^{-9.2\beta}$ \\
4 & 7 & $e^{-1.0\beta}$ \\
5 & 6 & $e^{-1.0\beta}$ \\
5 & 8 & $e^{-0.5\beta}$ \\
5 & 9 & $e^{-0.8\beta}$ \\
6 & 9 & $e^{-1.0\beta}$ \\
7 & 8 & $e^{-1.1\beta}$ \\
7 & 9 & $e^{-0.9\beta}$ \\
8 & 9 & $e^{-0.7\beta}$ \\
\end{tabular}
\end{center}
When we compute the edge weights using the value of $\beta$ that maximizes modularity, we obtain a graph that is optimally structured for community detection. In fact, we can see that edges with high multipliers in front of $\beta$, such as $(1, 7)$, $(1, 9)$, and $(2, 7)$ end up having weight very close to 0. In our time-aware model, the multipliers in front of $\beta_a$ are time differences between co-actions.

\begin{figure*}[h]
  \centering
  \includegraphics[width=\textwidth]{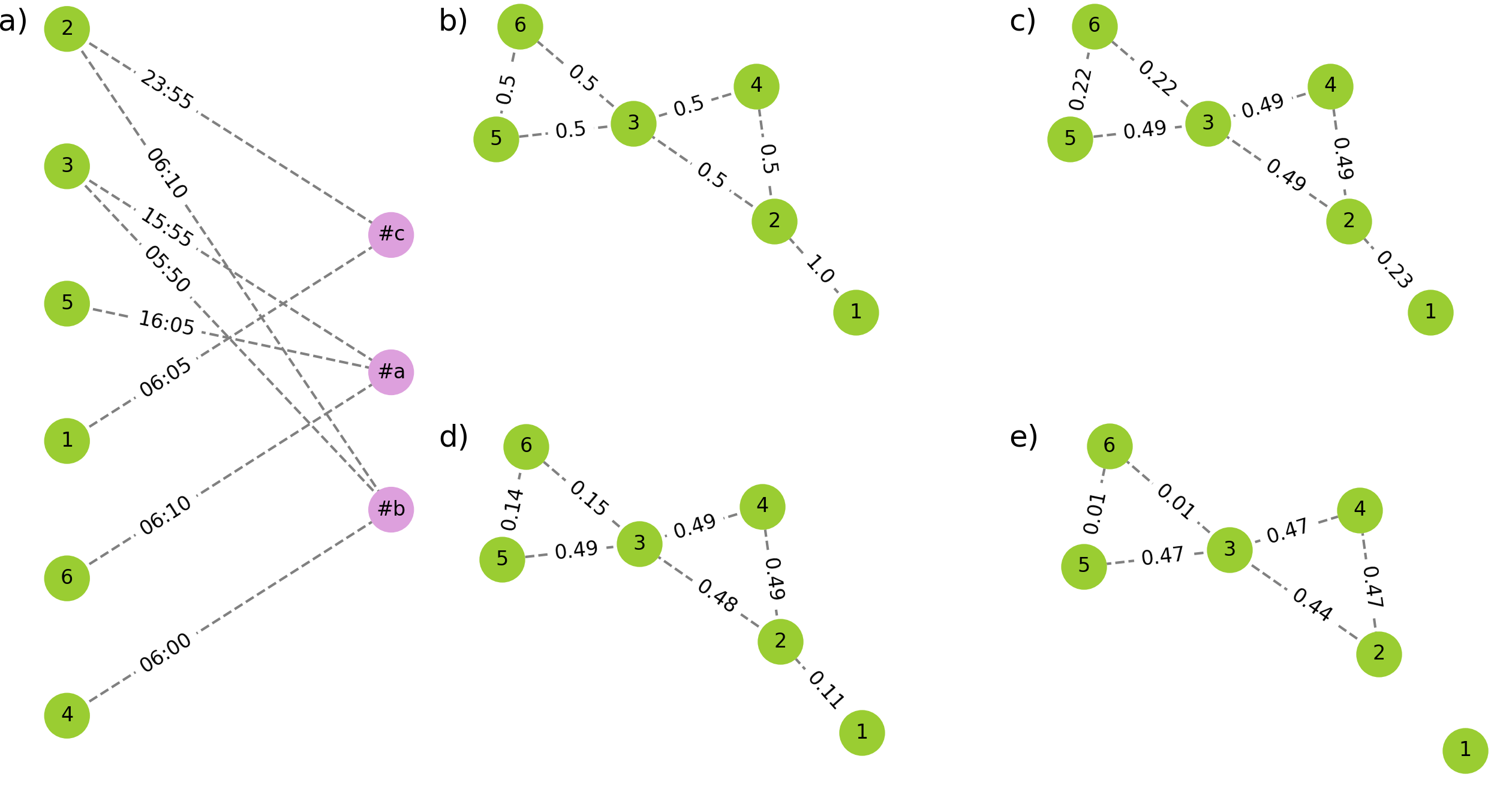}
  \caption{A minimal example illustrates the difference between the collaboration model proposed in~\cite{newman2001scientific} and our proposed time-aware model. Panel \textbf{a} shows user behavior over one day as a bipartite graph with time-stamped edges. Users 1, 2, 3, 4, 5, and 6 share hashtags \textit{\#a, \#b, \#c}. In Panel \textbf{b}, the same bipartite graph is folded into a collaboration network using the node-normalized model~\cite{newman2001scientific}, which does not consider the time dimension. When actions are not repeated, this corresponds to our proposed model with $\beta = 0$. As $\beta$ increases ($\beta = 2$ in Panel \textbf{c}, $\beta = 3$ in Panel \textbf{d}, and $\beta = 10$ in Panel \textbf{e}, with $\Delta t$ measured in minutes), co-actions with larger time differences are weighted less, resulting in a sparser network that only captures highly synchronized behavior.}
  \label{fig:time-aware-model}
\end{figure*}

\section{Performance of Time-aware Model on Synthetic Datasets}
\label{app:sim-results}

Synthetic datasets are generated by simulating 6 "inauthentic" users with a total of 15 to 20 timestamped activities, and 40 authentic users with a total of 1000 randomly timestamped activities distributed over 7 days. Random timestamps follow a uniform distribution that does not depend on the user. Timestamps for inauthentic users have variable rates over time intervals (average of 1.3 activities per minute in the active windows in Simulation 1, average of 1 activity per minutes in each of the active windows in Simulation 2, and averages of 1, 1.2, 1.3 and 1.5 activities per minutes in Simulation 3). A set of 20 different actions belonging to 3 different action types are randomly assigned to users, by setting a bimodal probability distribution over the frequency of the actions to ensure some actions are more common than others. The bimodal distribution is a mixture of two normal distributions $\mathcal{N}(3, 1)$ and $\mathcal{N}(12,1)$ and weight $0.6$. "Inauthentic" users perform on average 2 distinct actions on each modality, one with $p=0.9$ and the other one with $p=0.1$.

Table~\ref{table:multiplex-simulation} reports F1*, Homogeneity, and Weighted Precision scores obtained by applying the time-aware model on our simulations. To enable fair comparison, we also introduce a random labeler baseline. The random labeler model takes as input the same number of non-coordinated and coordinated users as we generate in our simulation, and assigns community labels to them entirely at random. Given a predefined number of clusters, the performance of the random labeler represents the expected performance if a model would produce the same number of clusters without capturing any meaningful signal from the data. In case of 2 clusters, the random labeler obtains an average Weighted Precision of $0.52$, average F1* of $0.67$, and average Homogeneity of $0.03$ over $1000$ repetitions. The corresponding $95\%$ confidence intervals are $(0.50, 0.58)$ for Weighted Precision, $(0.63,0.76)$ for F1*, and $(0.00, 0.15)$ for Homogeneity. In case of 3 clusters, the random labeler obtains an average Weighted Precision of $0.36$, average F1* of $0.56$, and average Homogeneity of $0.07$ over $1000$ repetitions. The corresponding $95\%$ confidence intervals are $(0.33, 0.43)$ for Weighted Precision, $(0.50,0.66)$ for  F1*, and $(0.00, 0.21)$ for Homogeneity. Even when inauthentic users are split into two separate clusters by the multiplex clustering, the results of our time-aware methods always outperform the random labeler on the same number of clusters.

\begin{table}[!h]
    \centering
    \begin{tabular}{c|c|c|c}
        Num. of Layers & F1* & Homog. & WP\\
        \hline
        1 (Sim. 1) & 1.00 & 1.00 & 1.00\\
        1 (Sim. 2) & 1.00 & 1.00 & 1.00\\
        1 (Sim. 3) & 0.67 & 1.00 & 1.00\\
        2 (Sim. 1 + Sim. 2) & 1.00 & 1.00 & 1.00\\
        2 (Sim. 1 + Sim. 3) & 1.00 & 1.00 & 1.00\\
        2 (Sim. 2 + Sim. 3) & 1.00 & 1.00 & 1.00\\
        3 (Sim. 1 + Sim. 2 + Sim. 3) & 1.00 & 1.00 & 1.00\\ 
    \end{tabular}
    \caption{Performance of time-aware model on simulated data.}
    \label{table:multiplex-simulation}
\end{table}

\section{Performance of Coordination Detection Approaches on Various Datasets}
\label{app:full-results}

We implement and evaluate a number of methods for coordinated behavior detection based on previous research. We provide their implementation as well as the implementation of our time-aware model at \url{https://github.com/letiziaia/time-aware-collaboration}. Numerical results for all selected metrics and all datasets are reported in Table~\ref{table:results_flipped_simple} and~\ref{table:results_h_wp}. 

The following paragraphs detail the specific parameters used for each method to ensure full reproducibility of the reported findings. Unless otherwise specified, the parameters used for each method are the same that are reported in the corresponding paper.

\textbf{Co-occurrence of hashtag sequences}~\citep{pacheco2021uncovering}: For each day in the dataset, we select the original posts only, and all users who have posted at least 5 different hashtags over the day. For each day, we build a network where two accounts are connected if they have shared an identical hashtag sequence. The daily networks are unweighted and no filtering is applied. The resulting connected components of each network are then interpreted as potentially coordinated groups of accounts. All users in any connected component are grouped together and considered to be labeled by the method as "coordinated". The method is then scored by comparing this list of users to the binary ground truth.  

\textbf{Rapid retweets}~\citep{pacheco2020unveiling, vishnuprasad2024tracking}: We build a directed network where two users are connected if one has reposted the other within a predefined time interval from the publication of the original post, and the weight of the edge is the number of times the promoter has been reposting content from the source account. We then remove self-loops and filter edges that have weight $>= 2$. We use 10 seconds~\citep{pacheco2020unveiling}, 30 seconds~\citep{vishnuprasad2024tracking}, and 60 seconds~\citep{vishnuprasad2024tracking} as time intervals. All nodes that have at least degree 1 are considered to be labeled by the method as "coordinated". The method is then scored by comparing this list of users to the binary ground truth.

\textbf{Ratclif-Obershelp similarity of tweets}~\citep{pacheco2020unveiling, vishnuprasad2024tracking}: We select all original posts and sort them by their publication timestamp. We then build a similarity network where two posts are connected if their distance in the timeline is at most 10 and if their Ratclif-Obershelp similarity score is $>= 0.7$. All nodes that have at least degree 1 are considered to be labeled by the method as "coordinated". The method is then scored by comparing this list of users to the binary ground truth.

\textbf{Cardinality of co-retweets}~\citep{linhares2022uncovering}: This method uses a time window of 7 days. For each time window, we build a co-retweet network such that two users are connected if they reposted the same post, and the weight of the edge is the number of repost they have in common. We then apply disparity filter with $alpha=0.05$, filter edges in each network based on neighborhood overlap with $k=0.05$, and discard all isolate nodes. All nodes that have at least degree 1 are considered to be labeled by the method as "coordinated". The method is then scored by comparing this list of users to the binary ground truth.

\textbf{AMDN-HAGE, AMDN-HAGE+KMEANS}~\citep{sharma2021identifying}: 
Prior to models execution, we preprocess datasets by constructing co-reposts and co-replies cascades, and filtering out users with fewer than $10$ activities within these cascades. For both models, the embedding size is set to $d=64$, the number of HAGE-components to $gmm_k=2$, the patience to $p=20$, and the learning rate to $lr=0.001$. For evaluation, a user is considered truly coordinated if he participates in at least one coordinated activity. Between the two resulting clusters, the cluster with the highest F1 score is designated as the predicted coordinated user group. Due to the initial filtering of users with low co-activity, some truly coordinated users are excluded from the models' clustering. To ensure a comprehensive performance assessment, these filtered-out, low-activity, true coordinated users are added to the false negative group. 

\textbf{Synchronized Action Framework} (hashtags, mentions, urls)~\citep{magelinski2022synchronized, ng2022online}: We consider all original posts and work over sliding time windows of size 5 minutes~\citep{magelinski2022synchronized, ng2022online}. We then consider all different hashtags in the dataset and build a network where two users are connected if they have shared the same hashtag within the time window. For each time window, the weight is the minimum of the activity count between the two users: if user A has shared hashtag \textit{foo} twice and user B three times within one time window, this time window contributes 2 to the edge weight~\citep{magelinski2022synchronized, ng2022online}. Time windows slide always to the following post in the timeline. We then apply Louvain community detection to the resulting weighted network~\citep{magelinski2022synchronized} and score the method by comparing each community of users to the binary ground truth. We report the result of the community with highest F1 score. The same pipeline is also applied to mentions, and to hashtag.
We also implement and score the variation presented by~\citet{ng2022online}. where the network is filtered by retaining only edges with weight $>= \text{ceil}(mean + std)$ before applying Louvain community detection. This variation is scored and reported in the same way as the original method in~\citet{magelinski2022synchronized}.

\textbf{BLOC} (Behavioral Language for Online Classification)~\citep{nwala2023language}: Each post is classified into action types (original post, reshare of another user, or self-reshare) and content types (text, hashtag, URL presence). These classifications are then encoded into BLOC strings using a symbolic alphabet and pause symbols. The implementation uses logarithmic time bins (pause up to one hour, one hour to one day, one day to one week, one week to one month, one month to one year, longer than a year), by considering as pause a time difference of more than a minute~\citep{nwala2023language}. We then extract character bigrams and compute the pairwise cosine similarity of the TF-IDF vectors for each users, but only retain similarity $\geq 0.98$~\citep{nwala2023language}. We finally remove singleton nodes and apply Louvain community detection to identify clusters of similar accounts~\citep{nwala2023language}.

\begin{table*}[!th]
\small
\centering

\renewcommand\arraystretch{0.3} % Reduce row spacing
\setlength\tabcolsep{2pt} % Reduce columns spacing
\setlength{\aboverulesep}{+0.2ex}
\setlength{\belowrulesep}{+0.2ex}

\def\rotationangle{90} % Define rotation angle
\def\rotationangledataset{0} % Define rotation angle
\def\methodfontsize{\scriptsize} % Define font size for methods
\def\datasetfontsize{\scriptsize} % Define font size for datasets (\small or \footnotesize)
\def\metricfontsize{\scriptsize} % Define font size for Precision, Recall, F1 - default is normal

\begin{tabular}{@{}ll|cccccccccccccccccccc@{}}
&  
& {\rotatebox{\rotationangle}{\methodfontsize \makecell{co-occurrence of hashtag sequences~\citep{pacheco2021uncovering}}} }
& {\rotatebox{\rotationangle}{\methodfontsize \makecell{rapid retweets (10 sec.)~\citep{pacheco2020unveiling} }} }
& {\rotatebox{\rotationangle}{\methodfontsize \makecell{rapid RT (30 sec.)~\citep{vishnuprasad2024tracking} }} }
& {\rotatebox{\rotationangle}{\methodfontsize \makecell{rapid RT (60 sec.)~\citep{vishnuprasad2024tracking} }} }
& {\rotatebox{\rotationangle}{\methodfontsize \makecell{cardinality of co-RT~\citep{linhares2022uncovering}}} }
& {\rotatebox{\rotationangle}{\methodfontsize \makecell{Ratclif-Obershelp\citep{vishnuprasad2024tracking, pacheco2020unveiling}}} }
& {\rotatebox{\rotationangle}{\methodfontsize \makecell{synchronized actions (hashtags)~\citep{magelinski2022synchronized}}} }
& {\rotatebox{\rotationangle}{\methodfontsize \makecell{synchronized actions + filtering (hashtags)~\citep{ng2022online}}} }
& {\rotatebox{\rotationangle}{\methodfontsize \makecell{synchronized actions (mentions)~\citep{magelinski2022synchronized}}} }
& {\rotatebox{\rotationangle}{\methodfontsize \makecell{synchronized actions + filtering (mentions)~\citep{ng2022online}}} }
& {\rotatebox{\rotationangle}{\methodfontsize \makecell{synchronized actions (urls)~\citep{magelinski2022synchronized}}} }
& {\rotatebox{\rotationangle}{\methodfontsize \makecell{synchronized actions + filtering (urls)~\citep{ng2022online}}} }
& {\rotatebox{\rotationangle}{\methodfontsize \makecell{AMDN-HAGE~\citep{sharma2021identifying}}} }
& {\rotatebox{\rotationangle}{\methodfontsize \makecell{AMDN-HAGE+KMEANS~\citep{sharma2021identifying}}} }
& {\rotatebox{\rotationangle}{\methodfontsize \makecell{BLOC~\citep{nwala2023language}}}  }
& {\rotatebox{\rotationangle}{\methodfontsize \makecell{Monoplex time-aware (hashtags)}} }
& {\rotatebox{\rotationangle}{\methodfontsize \makecell{Monoplex time-aware (retweets)}} }
& {\rotatebox{\rotationangle}{\methodfontsize \makecell{Monoplex time-aware (mentions)}} }
& {\rotatebox{\rotationangle}{\methodfontsize \makecell{Monoplex time-aware (urls)}} }
& {\rotatebox{\rotationangle}{\methodfontsize \makecell{Multiplex time-aware (4 layers)}}} \\
\midrule

\multirow{3}{*}{\rotatebox{\rotationangledataset}{\datasetfontsize Armenia}} & \metricfontsize Precision* & 0.60 & 0.00 & 0.00 & 0.00 & 0.00 & 0.06 & 0.86 & \textbf{1.00} & \textbf{1.00} & \textbf{1.00} & \textbf{1.00} & \textbf{1.00} & 0.16 & 0.16 & \textbf{1.00} & 0.45 & \textbf{1.00} & 0.75 & 0.50 & \textbf{1.00} \\
 & \metricfontsize Recall* & 0.48 & 0.00 & 0.00 & 0.00 & 0.00 & \textbf{0.77} & 0.58 & 0.03 & 0.10 & 0.03 & 0.03 & 0.03 & 0.29 & 0.29 & 0.13 & 0.61 & 0.55 & 0.39 & 0.03 & 0.55 \\
 & \metricfontsize F1* & 0.54 & 0.00 & 0.00 & 0.00 & 0.00 & 0.12 & 0.69 & 0.06 & 0.18 & 0.06 & 0.06 & 0.06 & 0.21 & 0.20 & 0.23 & 0.52 & \textbf{0.71} & 0.51 & 0.06 & 0.71 \\
\hline
\multirow{3}{*}{\rotatebox{\rotationangledataset}{\datasetfontsize Bangladesh}} & \metricfontsize Precision* & 0.00 & 0.00 & 0.00 & \textbf{1.00} & 0.00 & 0.04 & 0.19 & \textbf{1.00} & \textbf{1.00} & \textbf{1.00} & 0.00 & 0.00 & 0.09 & 0.08 & \textbf{1.00} & 0.07 & 0.20 & 0.09 & 0.50 & 0.50 \\
 & \metricfontsize Recall* & 0.00 & 0.00 & 0.00 & 0.18 & 0.00 & \textbf{0.73} & 0.27 & 0.09 & 0.18 & 0.09 & 0.00 & 0.00 & 0.18 & 0.09 & 0.27 & 0.45 & 0.09 & 0.27 & 0.09 & 0.18 \\
 & \metricfontsize F1* & 0.00 & 0.00 & 0.00 & 0.31 & 0.00 & 0.07 & 0.22 & 0.17 & 0.31 & 0.17 & 0.00 & 0.00 & 0.12 & 0.08 & \textbf{0.43} & 0.12 & 0.13 & 0.14 & 0.15 & 0.27 \\
\hline
\multirow{3}{*}{\rotatebox{\rotationangledataset}{\datasetfontsize Catalonia}} & \metricfontsize Precision* & 0.00 & 0.00 & 0.00 & 0.62 & 0.00 & 0.07 & 0.74 & 0.83 & 0.70 & \textbf{1.00} & 0.00 & 0.00 & 0.05 & 0.04 & \textbf{1.00} & 0.13 & 0.52 & 0.13 & 0.00 & 0.81 \\
 & \metricfontsize Recall* & 0.00 & 0.00 & 0.00 & 0.11 & 0.00 & 0.32 & 0.18 & 0.07 & 0.28 & 0.07 & 0.00 & 0.00 & 0.41 & 0.39 & 0.28 & 0.20 & \textbf{0.87} & 0.38 & 0.00 & 0.38 \\
 & \metricfontsize F1* & 0.00 & 0.00 & 0.00 & 0.18 & 0.00 & 0.12 & 0.29 & 0.12 & 0.40 & 0.12 & 0.00 & 0.00 & 0.08 & 0.08 & 0.43 & 0.16 & \textbf{0.65} & 0.20 & 0.00 & 0.52 \\
\hline
\multirow{3}{*}{\rotatebox{\rotationangledataset}{\datasetfontsize China 1}} & \metricfontsize Precision* & 0.05 & 0.00 & 0.08 & 0.26 & 0.00 & 0.04 & 0.86 & \textbf{1.00} & 0.85 & \textbf{1.00} & \textbf{1.00} & \textbf{1.00} & 0.00 & 0.00 & \textbf{1.00} & 0.62 & 0.97 & 0.64 & \textbf{1.00} & 0.98 \\
 & \metricfontsize Recall* & 0.01 & 0.00 & 0.01 & 0.06 & 0.00 & \textbf{0.69} & 0.53 & 0.00 & 0.27 & 0.01 & 0.02 & 0.00 & 0.00 & 0.00 & 0.20 & 0.58 & 0.10 & 0.12 & 0.04 & 0.06 \\
 & \metricfontsize F1* & 0.02 & 0.00 & 0.02 & 0.10 & 0.00 & 0.08 & \textbf{0.66} & 0.00 & 0.42 & 0.01 & 0.04 & 0.00 & 0.00 & 0.00 & 0.34 & 0.60 & 0.18 & 0.20 & 0.07 & 0.11 \\
\hline
\multirow{3}{*}{\rotatebox{\rotationangledataset}{\datasetfontsize China 2}} & \metricfontsize Precision* & 0.02 & 0.29 & 0.12 & 0.12 & 0.00 & 0.01 & 0.90 & \textbf{1.00} & 0.17 & \textbf{1.00} & 0.29 & \textbf{1.00} & 0.00 & 0.00 & \textbf{1.00} & 0.45 & 0.60 & 0.11 & 0.11 & \textbf{1.00} \\
 & \metricfontsize Recall* & 0.03 & 0.01 & 0.05 & 0.08 & 0.00 & \textbf{0.71} & 0.05 & 0.01 & 0.14 & 0.02 & 0.04 & 0.02 & 0.00 & 0.00 & 0.06 & 0.07 & 0.06 & 0.13 & 0.07 & 0.03 \\
 & \metricfontsize F1* & 0.03 & 0.02 & 0.07 & 0.10 & 0.00 & 0.02 & 0.09 & 0.01 & \textbf{0.15} & 0.03 & 0.07 & 0.03 & 0.00 & 0.00 & 0.11 & 0.13 & 0.11 & 0.12 & 0.09 & 0.05 \\
\hline
\multirow{3}{*}{\rotatebox{\rotationangledataset}{\datasetfontsize Cuba}} & \metricfontsize Precision* & 0.17 & 0.87 & 0.77 & 0.75 & \textbf{1.00} & 0.03 & 0.81 & 0.95 & 0.89 & \textbf{1.00} & \textbf{1.00} & \textbf{1.00} & 0.68 & 0.68 & 0.26 & 0.00 & 0.00 & 0.60 & 0.30 & 0.98 \\
 & \metricfontsize Recall* & 0.09 & 0.12 & 0.37 & 0.59 & 0.07 & 0.71 & \textbf{0.79} & 0.04 & 0.57 & 0.13 & 0.03 & 0.00 & 0.72 & 0.72 & 0.67 & 0.00 & 0.00 & 0.71 & 0.04 & 0.41 \\
 & \metricfontsize F1* & 0.12 & 0.21 & 0.50 & 0.66 & 0.13 & 0.07 & \textbf{0.80} & 0.07 & 0.69 & 0.24 & 0.06 & 0.00 & 0.70 & 0.70 & 0.38 & 0.00 & 0.00 & 0.65 & 0.07 & 0.57 \\
\hline
\multirow{3}{*}{\rotatebox{\rotationangledataset}{\datasetfontsize Ecuador}} & \metricfontsize Precision* & 0.55 & 0.94 & 0.57 & 0.53 & 0.90 & 0.13 & 0.87 & \textbf{1.00} & 0.96 & \textbf{1.00} & \textbf{1.00} & \textbf{1.00} & 0.15 & 0.15 & 0.04 & 0.66 & 0.62 & 0.77 & \textbf{1.00} & \textbf{1.00} \\
 & \metricfontsize Recall* & 0.11 & 0.08 & 0.10 & 0.13 & 0.02 & \textbf{0.83} & 0.46 & 0.00 & 0.30 & 0.11 & 0.00 & 0.00 & 0.45 & 0.45 & 0.24 & 0.50 & 0.20 & 0.48 & 0.00 & 0.10 \\
 & \metricfontsize F1* & 0.19 & 0.15 & 0.17 & 0.21 & 0.05 & 0.23 & 0.60 & 0.00 & 0.46 & 0.20 & 0.01 & 0.01 & 0.23 & 0.23 & 0.06 & 0.57 & 0.30 & \textbf{0.59} & 0.01 & 0.18 \\
\hline
\multirow{3}{*}{\rotatebox{\rotationangledataset}{\datasetfontsize Egypt}} & \metricfontsize Precision* & \textbf{1.00} & \textbf{1.00} & \textbf{1.00} & \textbf{1.00} & 0.00 & 0.64 & \textbf{1.00} & \textbf{1.00} & \textbf{1.00} & \textbf{1.00} & 0.00 & 0.00 & 0.92 & 0.92 & 0.94 & 0.76 & \textbf{1.00} & 0.95 & 0.00 & \textbf{1.00} \\
 & \metricfontsize Recall* & 0.09 & 0.02 & 0.04 & 0.19 & 0.00 & \textbf{0.80} & 0.28 & 0.01 & 0.10 & 0.06 & 0.00 & 0.00 & 0.59 & 0.59 & 0.33 & 0.48 & 0.35 & 0.17 & 0.00 & 0.04 \\
 & \metricfontsize F1* & 0.16 & 0.04 & 0.07 & 0.32 & 0.00 & 0.71 & 0.44 & 0.02 & 0.17 & 0.12 & 0.00 & 0.00 & \textbf{0.72} & \textbf{0.72} & 0.49 & 0.59 & 0.51 & 0.30 & 0.00 & 0.08 \\
\hline
\multirow{3}{*}{\rotatebox{\rotationangledataset}{\datasetfontsize Ghana}} & \metricfontsize Precision* & \textbf{1.00} & \textbf{1.00}& 0.82 & 0.89 & 0.00 & 0.16 & 0.86 & \textbf{1.00} & \textbf{1.00} & \textbf{1.00} & 0.00 & 0.00 & 0.26 & 0.26 &\textbf{1.00} & 0.19 & 0.67 & 0.44 & 0.00 & \textbf{1.00} \\
 & \metricfontsize Recall* & 0.10 & 0.12 & 0.15 & 0.28 & 0.00 & \textbf{0.73} & 0.42 & 0.05 & 0.20 & 0.03 & 0.00 & 0.00 & 0.35 & 0.35 & 0.01 & 0.37 & 0.37 & 0.25 & 0.00 & 0.45 \\
 & \metricfontsize F1* & 0.18 & 0.21 & 0.25 & 0.43 & 0.00 & 0.27 & 0.56 & 0.10 & 0.33 & 0.06 & 0.00 & 0.00 & 0.30 & 0.30 & 0.18 & 0.25 & 0.47 & 0.32 & 0.00 & \textbf{0.62} \\
\hline
\multirow{3}{*}{\rotatebox{\rotationangledataset}{\datasetfontsize Iran 1}} & \metricfontsize Precision* & 0.57 & 0.92 & 0.89 & 0.94 & \textbf{1.00} & 0.21 & 0.74 & \textbf{1.00} & \textbf{1.00} & \textbf{1.00} & \textbf{1.00} & 0.00 & 0.00 & 0.00 &\textbf{1.00} & 0.49 & \textbf{1.00} & 0.98 & \textbf{1.00} & \textbf{1.00} \\
 & \metricfontsize Recall* & 0.09 & 0.07 & 0.10 & 0.19 & 0.00 & \textbf{0.53} & 0.19 & 0.01 & 0.09 & 0.01 & 0.00 & 0.00 & 0.00 & 0.00 & 0.18 & 0.29 & 0.28 & 0.26 & 0.00 & 0.03 \\
 & \metricfontsize F1* & 0.15 & 0.12 & 0.18 & 0.31 & 0.01 & 0.30 & 0.31 & 0.02 & 0.16 & 0.02 & 0.01 & 0.00 & 0.00 & 0.00 & 0.30 & 0.37 & 0.44 & \textbf{0.41} & 0.01 & 0.06 \\
\hline
\multirow{3}{*}{\rotatebox{\rotationangledataset}{\datasetfontsize Iran 2}} & \metricfontsize Precision* & 0.11 & \textbf{1.00} & 0.86 & 0.81 & 0.00 & 0.08 & \textbf{1.00} & \textbf{1.00} & \textbf{1.00} & \textbf{1.00} & \textbf{1.00} & 0.00 & 0.00 & 0.00 &\textbf{1.00}& \textbf{1.00} & \textbf{1.00} & 0.96 & 0.00 & \textbf{1.00} \\
 & \metricfontsize Recall* & 0.04 & 0.03 & 0.08 & 0.12 & 0.00 & \textbf{0.43} & 0.14 & 0.00 & 0.07 & 0.02 & 0.00 & 0.00 & 0.00 & 0.00 & 0.19 & 0.13 & 0.21 & 0.13 & 0.00 & 0.06 \\
 & \metricfontsize F1* & 0.05 & 0.05 & 0.15 & 0.21 & 0.00 & 0.13 & 0.24 & 0.01 & 0.14 & 0.03 & 0.01 & 0.00 & 0.00 & 0.00 & 0.32 & 0.23 & \textbf{0.34} & 0.23 & 0.00 & 0.11 \\
\hline
\multirow{3}{*}{\rotatebox{\rotationangledataset}{\datasetfontsize Iran 3}} & \metricfontsize Precision* & 0.00 & \textbf{1.00} & 0.88 & 0.94 & 0.00 & 0.14 & 0.95 & \textbf{1.00} & 0.97 & \textbf{1.00} & \textbf{1.00} & 0.00 & 0.00 & 0.00 & 0.70 & 0.36 & \textbf{1.00} & 0.33 & 0.00 & \textbf{1.00} \\
 & \metricfontsize Recall* & 0.00 & 0.03 & 0.07 & 0.16 & 0.00 & \textbf{0.45} & 0.34 & 0.00 & 0.16 & 0.02 & 0.00 & 0.00 & 0.00 & 0.00 & 0.34 & 0.41 & 0.24 & 0.25 & 0.00 & 0.09 \\
 & \metricfontsize F1* & 0.00 & 0.06 & 0.13 & 0.28 & 0.00 & 0.21 & \textbf{0.50} & 0.01 & 0.28 & 0.04 & 0.01 & 0.00 & 0.00 & 0.00 & 0.46 & 0.38 & 0.39 & 0.29 & 0.00 & 0.17 \\
\hline
\multirow{3}{*}{\rotatebox{\rotationangledataset}{\datasetfontsize Iran 4}} & \metricfontsize Precision* & 0.35 & \textbf{1.00} & 0.90 & 0.94 & 0.00 & 0.44 & 0.99 & \textbf{1.00} & \textbf{1.00} & \textbf{1.00} & \textbf{1.00} & \textbf{1.00} & 0.00 & 0.00 & \textbf{1.00} & 0.97 & 0.99 & \textbf{1.00} & 0.00 & \textbf{1.00} \\
 & \metricfontsize Recall* & 0.01 & 0.00 & 0.01 & 0.03 & 0.00 & 0.47 & 0.38 & 0.00 & 0.11 & 0.00 & 0.00 & 0.00 & 0.00 & 0.00 & 0.39 & \textbf{0.63} & 0.10 & 0.31 & 0.00 & 0.01 \\
 & \metricfontsize F1* & 0.02 & 0.01 & 0.02 & 0.06 & 0.00 & 0.45 & 0.55 & 0.01 & 0.19 & 0.01 & 0.00 & 0.00 & 0.00 & 0.00 & 0.56 & \textbf{0.76} & 0.18 & 0.47 & 0.00 & 0.02 \\
\hline
\multirow{3}{*}{\rotatebox{\rotationangledataset}{\datasetfontsize Iran 5}} & \metricfontsize Precision* & 0.00 & 0.00 & 0.00 & 0.50 & 0.00 & 0.11 & \textbf{1.00} & \textbf{1.00} & \textbf{1.00} & \textbf{1.00} & 0.00 & 0.00 & 0.00 & 0.00 & \textbf{1.00} & \textbf{1.00} & 0.96 & 0.49 & 0.00 & 0.96 \\
 & \metricfontsize Recall* & 0.00 & 0.00 & 0.00 & 0.02 & 0.00 & 0.23 & 0.25 & 0.01 & 0.08 & 0.04 & 0.00 & 0.00 & 0.00 & 0.00 & 0.54 & \textbf{0.77} & 0.22 & 0.22 & 0.00 & 0.21 \\
 & \metricfontsize F1* & 0.00 & 0.00 & 0.00 & 0.04 & 0.00 & 0.15 & 0.40 & 0.02 & 0.14 & 0.07 & 0.00 & 0.00 & 0.00 & 0.00 & 0.70 & \textbf{0.87} & 0.36 & 0.30 & 0.00 & 0.35 \\
\hline
\multirow{3}{*}{\rotatebox{\rotationangledataset}{\datasetfontsize Iran 6}} & \metricfontsize Precision* & 0.32 & \textbf{1.00} & 0.79 & 0.74 & 0.00 & 0.05 & \textbf{1.00} & \textbf{1.00} & 0.72 & \textbf{1.00} & \textbf{1.00} & \textbf{1.00} & 0.00 & 0.00 & 0.04 & 0.87 & 0.15 & 0.95 & 0.00 & 0.95 \\
 & \metricfontsize Recall* & 0.09 & 0.05 & 0.11 & 0.16 & 0.00 & \textbf{0.68} & 0.36 & 0.03 & 0.23 & 0.02 & 0.00 & 0.00 & 0.00 & 0.00 & 0.32 & 0.36 & 0.22 & 0.17 & 0.00 & 0.09 \\
 & \metricfontsize F1* & 0.14 & 0.09 & 0.19 & 0.27 & 0.00 & 0.09 & \textbf{0.53} & 0.06 & 0.35 & 0.04 & 0.01 & 0.01 & 0.00 & 0.00 & 0.07 & 0.51 & 0.18 & 0.28 & 0.00 & 0.17 \\
\hline
\multirow{3}{*}{\rotatebox{\rotationangledataset}{\datasetfontsize Qatar}} & \metricfontsize Precision* & 0.00 & 0.00 & 0.00 & 0.06 & 0.00 & 0.00 & 0.18 & \textbf{1.00} & 0.88 & \textbf{1.00} & \textbf{1.00} & 0.00 & 0.01 & 0.01 & 0.75 & 0.06 & 0.04 & 0.14 & 0.00 & \textbf{1.00} \\
 & \metricfontsize Recall* & 0.00 & 0.00 & 0.00 & 0.14 & 0.00 & 0.52 & 0.14 & 0.03 & 0.24 & 0.03 & 0.03 & 0.00 & 0.62 & \textbf{0.66} & 0.10 & 0.07 & 0.38 & 0.17 & 0.00 & 0.14 \\
 & \metricfontsize F1* & 0.00 & 0.00 & 0.00 & 0.08 & 0.00 & 0.01 & 0.16 & 0.07 & \textbf{0.38} & 0.07 & 0.07 & 0.00 & 0.02 & 0.02 & 0.18 & 0.06 & 0.07 & 0.15 & 0.00 & 0.24 \\
\hline
\multirow{3}{*}{\rotatebox{\rotationangledataset}{\datasetfontsize Russia 1}} & \metricfontsize Precision* & 0.25 & 0.75 & 0.50 & 0.59 & 0.87 & 0.18 & 0.98 & \textbf{1.00} & \textbf{1.00} & \textbf{1.00} & \textbf{1.00} & \textbf{1.00} & 0.00 & 0.00 & 0.78 & 0.98 & \textbf{1.00} & \textbf{1.00} & \textbf{1.00} & \textbf{1.00} \\
 & \metricfontsize Recall* & 0.09 & 0.01 & 0.04 & 0.12 & 0.01 & 0.64 & \textbf{0.70} & 0.00 & 0.27 & 0.00 & 0.00 & 0.00 & 0.00 & 0.00 & 0.24 & 0.35 & 0.31 & 0.18 & 0.02 & 0.10 \\
 & \metricfontsize F1* & 0.13 & 0.01 & 0.08 & 0.20 & 0.02 & 0.28 & \textbf{0.82} & 0.00 & 0.42 & 0.00 & 0.00 & 0.00 & 0.00 & 0.00 & 0.37 & 0.52 & 0.47 & 0.31 & 0.04 & 0.18 \\
\hline
\multirow{3}{*}{\rotatebox{\rotationangledataset}{\datasetfontsize Russia 2}} & \metricfontsize Precision* & 0.10 & \textbf{1.00} & 0.67 & 0.82 & 0.00 & 0.26 & \textbf{1.00} & \textbf{1.00} & \textbf{1.00} & \textbf{1.00} & 0.00 & 0.00 & 0.00 & 0.00 & 0.64 & \textbf{1.00} & \textbf{1.00} & \textbf{1.00} & \textbf{1.00} & \textbf{1.00} \\
 & \metricfontsize Recall* & 0.01 & 0.01 & 0.01 & 0.02 & 0.00 & \textbf{0.58} & 0.21 & 0.00 & 0.02 & 0.01 & 0.00 & 0.00 & 0.00 & 0.00 & 0.38 & 0.20 & 0.20 & 0.04 & 0.01 & 0.02 \\
 & \metricfontsize F1* & 0.02 & 0.01 & 0.02 & 0.05 & 0.00 & 0.36 & 0.35 & 0.01 & 0.05 & 0.01 & 0.00 & 0.00 & 0.00 & 0.00 & \textbf{0.48} & 0.33 & 0.34 & 0.08 & 0.03 & 0.04 \\
\hline
\multirow{3}{*}{\rotatebox{\rotationangledataset}{\datasetfontsize Russia 3}} & \metricfontsize Precision* & 0.00 & 0.00 & 0.00 & 0.00 & 0.00 & 0.01 & \textbf{1.00} & \textbf{1.00} & \textbf{1.00} & \textbf{1.00} & 0.00 & 0.00 & 0.00 & 0.00 & 0.00 & 0.17 & 0.50 & 0.17 & 0.00 & 0.50 \\
 & \metricfontsize Recall* & 0.00 & 0.00 & 0.00 & 0.00 & 0.00 & 0.40 & 0.20 & 0.20 & 0.20 & 0.20 & 0.00 & 0.00 & 0.00 & 0.00 & 0.00 & 0.20 & \textbf{0.80} & 0.20 & 0.00 & 0.60 \\
 & \metricfontsize F1* & 0.00 & 0.00 & 0.00 & 0.00 & 0.00 & 0.03 & 0.33 & 0.33 & 0.33 & 0.33 & 0.00 & 0.00 & 0.00 & 0.00 & 0.00 & 0.18 & \textbf{0.62} & 0.18 & 0.00 & 0.55 \\
\hline
\multirow{3}{*}{\rotatebox{\rotationangledataset}{\datasetfontsize Russia 4}} & \metricfontsize Precision* & 0.00 & \textbf{1.00} & 0.03 & 0.02 & 0.00 & 0.00 & \textbf{1.00} & \textbf{1.00} & \textbf{1.00} & \textbf{1.00} & \textbf{1.00} & \textbf{1.00} & 0.00 & 0.00 & 0.00 & 0.08 & 0.67 & 0.15 & 0.50 & 0.38 \\
 & \metricfontsize Recall* & 0.00 & 0.08 & 0.12 & 0.17 & 0.00 & \textbf{0.75} & 0.04 & 0.04 & 0.08 & 0.04 & 0.04 & 0.04 & 0.00 & 0.00 & 0.04 & 0.04 & 0.08 & 0.08 & 0.04 & 0.12 \\
 & \metricfontsize F1* & 0.00 & 0.15 & 0.05 & 0.04 & 0.00 & 0.00 & 0.08 & 0.08 & 0.15 & 0.08 & 0.08 & 0.08 & 0.00 & 0.00 & 0.00 & 0.06 & 0.15 & 0.11 & 0.08 & \textbf{0.19} \\
\hline
\multirow{3}{*}{\rotatebox{\rotationangledataset}{\datasetfontsize Russia 5}} & \metricfontsize Precision* & 0.19 & 0.00 & 0.00 & 0.06 & 0.00 & 0.01 & 0.12 & \textbf{1.00} & \textbf{1.00} & \textbf{1.00} & 0.00 & 0.00 & 0.00 & 0.00 & \textbf{1.00} & 0.06 & 0.33 & 0.09 & 0.33 & \textbf{1.00} \\
 & \metricfontsize Recall* & 0.12 & 0.00 & 0.00 & 0.06 & 0.00 & \textbf{0.51} & 0.27 & 0.02 & 0.06 & 0.02 & 0.00 & 0.00 & 0.00 & 0.00 & 0.04 & 0.24 & 0.25 & 0.04 & 0.02 & 0.22 \\
 & \metricfontsize F1* & 0.14 & 0.00 & 0.00 & 0.06 & 0.00 & 0.02 & 0.17 & 0.04 & 0.11 & 0.04 & 0.00 & 0.00 & 0.00 & 0.00 & 0.08 & 0.10 & 0.29 & 0.05 & 0.04 & \textbf{0.35} \\
\hline
\multirow{3}{*}{\rotatebox{\rotationangledataset}{\datasetfontsize Spain}} & \metricfontsize Precision* & 0.83 & 0.90 & 0.91 & 0.93 & 0.00 & 0.52 & 0.93 & 0.98 & 0.88 & \textbf{1.00} & 0.00 & 0.00 & 0.34 & 0.34 & 0.87 & 0.80 & 0.72 & 0.49 & 0.00 & \textbf{1.00} \\
 & \metricfontsize Recall* & 0.07 & 0.09 & 0.25 & 0.30 & 0.00 & \textbf{0.78} & 0.58 & 0.20 & 0.52 & 0.15 & 0.00 & 0.00 & 0.44 & 0.44 & 0.72 & 0.58 & 0.53 & 0.55 & 0.00 & 0.43 \\
 & \metricfontsize F1* & 0.13 & 0.16 & 0.39 & 0.45 & 0.00 & 0.62 & 0.71 & 0.33 & 0.66 & 0.26 & 0.00 & 0.00 & 0.38 & 0.38 & \textbf{0.79} & 0.67 & 0.61 & 0.52 & 0.00 & 0.60 \\
\hline
\multirow{3}{*}{\rotatebox{\rotationangledataset}{\datasetfontsize Thailand}} & \metricfontsize Precision* & 0.00 & 0.00 & 0.00 & 0.20 & 0.00 & 0.46 & \textbf{1.00} & \textbf{1.00} & 0.95 & \textbf{1.00} & 0.00 & 0.00 & 0.45 & 0.11 & 0.41 & 0.54 & 0.95 & 0.54 & 0.00 & \textbf{1.00} \\
 & \metricfontsize Recall* & 0.00 & 0.00 & 0.00 & 0.00 & 0.00 & \textbf{0.46} & 0.05 & 0.00 & 0.12 & 0.02 & 0.00 & 0.00 & 0.27 & 0.34 & 0.17 & 0.20 & 0.53 & 0.17 & 0.00 & 0.16 \\
 & \metricfontsize F1* & 0.00 & 0.00 & 0.00 & 0.01 & 0.00 & \textbf{0.46} & 0.10 & 0.00 & 0.21 & 0.04 & 0.00 & 0.00 & 0.34 & 0.17 & 0.24 & 0.29 & 0.68 & 0.26 & 0.00 & 0.28 \\
\hline
\multirow{3}{*}{\rotatebox{\rotationangledataset}{\datasetfontsize Uae}} & \metricfontsize Precision* & 0.99 & \textbf{1.00} & 0.91 & 0.93 & \textbf{1.00} & 0.58 & \textbf{1.00} & \textbf{1.00} & \textbf{1.00} & \textbf{1.00} & \textbf{1.00} & \textbf{1.00} & 0.68 & 0.68 & 0.88 & 0.96 & \textbf{1.00} & 0.95 & \textbf{1.00} & \textbf{1.00} \\
 & \metricfontsize Recall* & 0.38 & 0.01 & 0.07 & 0.16 & 0.00 & \textbf{0.78} & 0.14 & 0.02 & 0.10 & 0.01 & 0.00 & 0.00 & 0.71 & 0.71 & 0.46 & 0.20 & 0.15 & 0.13 & 0.00 & 0.03 \\
 & \metricfontsize F1* & 0.55 & 0.02 & 0.12 & 0.27 & 0.00 & 0.67 & 0.24 & 0.03 & 0.18 & 0.01 & 0.00 & 0.00 & \textbf{0.70} & \textbf{0.70} & 0.60 & 0.33 & 0.26 & 0.22 & 0.00 & 0.06 \\
\hline
\multirow{3}{*}{\rotatebox{\rotationangledataset}{\datasetfontsize Venezuela 1}} & \metricfontsize Precision* & 0.25 & 0.00 & 0.00 & 0.00 & 0.00 & 0.31 & 0.99 & \textbf{1.00} & \textbf{1.00} & \textbf{1.00} & \textbf{1.00} & 0.00 & 0.00 & 0.00 & 0.71 & 0.28 & \textbf{1.00} & 0.06 & \textbf{1.00} & \textbf{1.00} \\
 & \metricfontsize Recall* & 0.00 & 0.00 & 0.00 & 0.00 & 0.00 & 0.73 & 0.23 & 0.10 & 0.13 & 0.00 & 0.01 & 0.00 & 0.00 & 0.00 & \textbf{0.76} & 0.54 & 0.03 & 0.12 & 0.01 & 0.07 \\
 & \metricfontsize F1* & 0.01 & 0.00 & 0.00 & 0.00 & 0.00 & 0.43 & 0.37 & 0.18 & 0.23 & 0.00 & 0.01 & 0.00 & 0.00 & 0.00 & \textbf{0.73} & 0.37 & 0.06 & 0.08 & 0.01 & 0.13 \\
\hline
\multirow{3}{*}{\rotatebox{\rotationangledataset}{\datasetfontsize Venezuela 2}} & \metricfontsize Precision* & 0.50 & 0.00 & 0.00 & 0.00 & 0.00 & 0.05 & 0.31 & \textbf{1.00} & 0.73 & \textbf{1.00} & \textbf{1.00} & 0.00 & 0.00 & 0.00 & 0.24 & 0.24 & \textbf{1.00} & 0.70 & \textbf{1.00} & 0.65 \\
 & \metricfontsize Recall* & 0.06 & 0.00 & 0.00 & 0.00 & 0.00 & \textbf{0.94} & 0.55 & 0.27 & 0.24 & 0.09 & 0.06 & 0.00 & 0.00 & 0.00 & 0.79 & 0.12 & 0.36 & 0.21 & 0.12 & 0.33 \\
 & \metricfontsize F1* & 0.11 & 0.00 & 0.00 & 0.00 & 0.00 & 0.09 & 0.40 & 0.43 & 0.36 & 0.17 & 0.11 & 0.00 & 0.00 & 0.00 & 0.36 & 0.16 & \textbf{0.53} & 0.33 & 0.22 & 0.44 \\

\bottomrule
\end{tabular}
\caption{Performance of Coordination Detection Approaches on Various Datasets.}
\label{table:results_flipped_simple}

\end{table*}

\newpage

\begin{table*}[!th]
\small
\centering

\renewcommand\arraystretch{0.3} % Reduce row spacing
\setlength\tabcolsep{2pt} % Reduce columns spacing
\setlength{\aboverulesep}{+0.2ex}
\setlength{\belowrulesep}{+0.2ex}

\def\rotationangle{90} % Define rotation angle
\def\rotationangledataset{0} % Define rotation angle
\def\methodfontsize{\scriptsize} % Define font size for methods
\def\datasetfontsize{\scriptsize} % Define font size for datasets (\small or \footnotesize)
\def\metricfontsize{\scriptsize} % Define font size for Precision, Recall, F1 - default is normal

\begin{tabular}{@{}ll|cccccccccccccccccccc@{}}
&  
& {\rotatebox{\rotationangle}{\methodfontsize \makecell{co-occurrence of hashtag sequences~\citep{pacheco2021uncovering}}} }
& {\rotatebox{\rotationangle}{\methodfontsize \makecell{rapid retweets (10 sec.)~\citep{pacheco2020unveiling} }} }
& {\rotatebox{\rotationangle}{\methodfontsize \makecell{rapid RT (30 sec.)~\citep{vishnuprasad2024tracking} }} }
& {\rotatebox{\rotationangle}{\methodfontsize \makecell{rapid RT (60 sec.)~\citep{vishnuprasad2024tracking} }} }
& {\rotatebox{\rotationangle}{\methodfontsize \makecell{cardinality of co-RT~\citep{linhares2022uncovering}}} }
& {\rotatebox{\rotationangle}{\methodfontsize \makecell{Ratclif-Obershelp\citep{vishnuprasad2024tracking, pacheco2020unveiling}}} }
& {\rotatebox{\rotationangle}{\methodfontsize \makecell{synchronized actions (hashtags)~\citep{magelinski2022synchronized}}} }
& {\rotatebox{\rotationangle}{\methodfontsize \makecell{synchronized actions + filtering (hashtags)~\citep{ng2022online}}} }
& {\rotatebox{\rotationangle}{\methodfontsize \makecell{synchronized actions (mentions)~\citep{magelinski2022synchronized}}} }
& {\rotatebox{\rotationangle}{\methodfontsize \makecell{synchronized actions + filtering (mentions)~\citep{ng2022online}}} }
& {\rotatebox{\rotationangle}{\methodfontsize \makecell{synchronized actions (urls)~\citep{magelinski2022synchronized}}} }
& {\rotatebox{\rotationangle}{\methodfontsize \makecell{synchronized actions + filtering (urls)~\citep{ng2022online}}} }
& {\rotatebox{\rotationangle}{\methodfontsize \makecell{AMDN-HAGE~\citep{sharma2021identifying}}} }
& {\rotatebox{\rotationangle}{\methodfontsize \makecell{AMDN-HAGE+KMEANS~\citep{sharma2021identifying}}} }
& {\rotatebox{\rotationangle}{\methodfontsize \makecell{BLOC~\citep{nwala2023language}}}  }
& {\rotatebox{\rotationangle}{\methodfontsize \makecell{Monoplex time-aware (hashtags)}} }
& {\rotatebox{\rotationangle}{\methodfontsize \makecell{Monoplex time-aware (retweets)}} }
& {\rotatebox{\rotationangle}{\methodfontsize \makecell{Monoplex time-aware (mentions)}} }
& {\rotatebox{\rotationangle}{\methodfontsize \makecell{Monoplex time-aware (urls)}} }
& {\rotatebox{\rotationangle}{\methodfontsize \makecell{Multiplex time-aware (4 layers)}}} \\
\midrule

\multirow{3}{*}{\rotatebox{\rotationangledataset}{\datasetfontsize Armenia}} & \metricfontsize Homog. & 0.31 & 0.00 & 0.00 & 0.00 & 0.00 & 0.14 & 0.82 & \textbf{1.00} & 0.99 & \textbf{1.00} & 0.98 & 0.98 & 0.80 & 0.69 & 0.63 & 0.41 & 0.94 & 0.52 & 0.14 & 0.91 \\
 & \metricfontsize WP & 0.29 & 0.00 & 0.02 & 0.02 & 0.00 & 0.05 & 0.69 & 0.00 & \textbf{0.92} & 0.00 & 0.25 & 0.25 & 0.14 & 0.14 & 0.45 & 0.30 & \textbf{0.92} & 0.47 & 0.29 & 0.80 \\
 & \metricfontsize NMI & 0.34 & 0.00 & 0.00 & 0.00 & 0.00 & 0.00 & 0.51 & 0.00 & 0.25 & 0.00 & 0.00 & 0.00 & 0.06 & 0.05 & 0.45 & 0.00 & \textbf{0.88} & 0.44 & 0.09 & 0.59 \\
\hline
\multirow{3}{*}{\rotatebox{\rotationangledataset}{\datasetfontsize Bangladesh}} & \metricfontsize Homog. & 0.00 & 0.00 & 0.00 & 0.15 & 0.00 & 0.11 & 0.74 & \textbf{1.00} & 0.96 & \textbf{1.00} & 0.00 & 0.00 & 0.67 & 0.65 & 0.69 & 0.22 & 0.40 & 0.25 & 0.00 & 0.86 \\
 & \metricfontsize WP & 0.00 & 0.00 & 0.00 & 0.19 & 0.00 & 0.03 & 0.20 & 0.00 & \textbf{0.83} & 0.00 & 0.00 & 0.00 & 0.09 & 0.06 & 0.54 & 0.05 & 0.17 & 0.10 & 0.50 & 0.45 \\
 & \metricfontsize NMI & 0.00 & 0.00 & 0.00 & 0.25 & 0.00 & 0.00 & 0.00 & 0.00 & 0.37 & 0.00 & 0.00 & 0.00 & 0.05 & 0.04 & \textbf{0.56} & 0.00 & 0.00 & 0.00 & 0.00 & 0.19 \\
\hline
\multirow{3}{*}{\rotatebox{\rotationangledataset}{\datasetfontsize Catalonia}} & \metricfontsize Homog. & 0.00 & 0.00 & 0.00 & 0.06 & 0.00 & 0.03 & 0.88 & 0.98 & 0.87 & \textbf{1.00} & 0.00 & 0.00 & 0.45 & 0.35 & 0.96 & 0.36 & 0.65 & 0.28 & 0.00 & 0.92 \\
 & \metricfontsize WP & 0.00 & 0.00 & 0.00 & 0.09 & 0.00 & 0.04 & 0.69 & 0.83 & 0.77 & \textbf{1.00} & 0.00 & 0.00 & 0.05 & 0.04 & 0.96 & 0.25 & 0.51 & 0.13 & 0.00 & 0.85 \\
 & \metricfontsize NMI & 0.00 & 0.00 & 0.00 & 0.10 & 0.00 & 0.00 & 0.40 & 0.19 & 0.65 & 0.31 & 0.00 & 0.00 & 0.03 & 0.02 & \textbf{0.94} & 0.20 & 0.49 & 0.00 & 0.00 & 0.62 \\
\hline
\multirow{3}{*}{\rotatebox{\rotationangledataset}{\datasetfontsize China 1}} & \metricfontsize Homog. & 0.00 & 0.00 & 0.00 & 0.02 & 0.00 & 0.08 & 0.87 & \textbf{1.00} & 0.93 & \textbf{1.00} & \textbf{1.00} & \textbf{1.00} & 0.32 & 0.30 & 0.55 & 0.67 & 0.66 & 0.59 & 0.67 & 0.97 \\
 & \metricfontsize WP & 0.02 & 0.02 & 0.02 & 0.03 & 0.00 & 0.03 & 0.81 & 0.26 & 0.85 & 0.90 & 0.88 & 0.25 & 0.00 & 0.00 & 0.56 & 0.55 & 0.58 & 0.44 & 0.67 & \textbf{0.92} \\
 & \metricfontsize NMI & 0.00 & 0.00 & 0.00 & 0.00 & 0.00 & 0.00 & 0.70 & 0.00 & \textbf{0.61} & 0.01 & 0.07 & 0.00 & 0.02 & 0.02 & 0.59 & 0.53 & 0.55 & 0.33 & 0.50 & 0.55 \\
\hline
\multirow{3}{*}{\rotatebox{\rotationangledataset}{\datasetfontsize China 2}} & \metricfontsize Homog. & 0.00 & 0.01 & 0.02 & 0.03 & 0.00 & 0.03 & 0.73 & 1.00 & 0.83 & \textbf{1.00} & 0.97 & \textbf{1.00} & 0.29 & 0.28 & 0.31 & 0.37 & 0.62 & 0.37 & 0.37 & 0.90 \\
 & \metricfontsize WP & 0.01 & 0.01 & 0.01 & 0.01 & 0.00 & 0.01 & 0.42 & 0.25 & 0.45 & 0.79 & 0.65 & \textbf{1.00} & 0.00 & 0.00 & 0.27 & 0.15 & 0.53 & 0.15 & 0.21 & 0.63 \\
 & \metricfontsize NMI & 0.00 & 0.00 & 0.00 & 0.00 & 0.00 & 0.00 & 0.30 & 0.00 & 0.25 & 0.05 & 0.13 & 0.03 & 0.02 & 0.02 & 0.34 & 0.09 & \textbf{0.48} & 0.10 & 0.17 & 0.34 \\
\hline
\multirow{3}{*}{\rotatebox{\rotationangledataset}{\datasetfontsize Cuba}} & \metricfontsize Homog. & 0.03 & 0.09 & 0.26 & 0.43 & 0.06 & 0.06 & 0.82 & \textbf{1.00} & 0.93 & \textbf{1.00} & \textbf{1.00} & 1.00 & 0.41 & 0.44 & 0.45 & 0.00 & 0.00 & 0.66 & 0.40 & 0.98 \\
 & \metricfontsize WP & 0.03 & 0.12 & 0.29 & 0.44 & 0.08 & 0.03 & 0.76 & 0.97 & 0.91 & \textbf{0.99} & 0.95 & 0.00 & 0.58 & 0.57 & 0.24 & 0.00 & 0.00 & 0.54 & 0.29 & 0.97 \\
 & \metricfontsize NMI & 0.00 & 0.15 & 0.34 & 0.47 & 0.10 & 0.00 & 0.70 & 0.09 & \textbf{0.79} & 0.24 & 0.06 & 0.00 & 0.41 & 0.43 & 0.05 & 0.00 & 0.00 & 0.52 & 0.00 & 0.62 \\
\hline
\multirow{3}{*}{\rotatebox{\rotationangledataset}{\datasetfontsize Ecuador}} & \metricfontsize Homog. & 0.06 & 0.06 & 0.05 & 0.06 & 0.02 & 0.20 & 0.79 & \textbf{1.00} & 0.89 & 0.99 & \textbf{1.00} & \textbf{1.00} & 0.30 & 0.19 & 0.03 & 0.56 & 0.65 & 0.65 & 0.73 & 0.99 \\
 & \metricfontsize WP & 0.09 & 0.11 & 0.09 & 0.10 & 0.06 & 0.11 & 0.69 & 0.00 & 0.85 & 0.96 & \textbf{1.00} & \textbf{1.00} & 0.17 & 0.17 & 0.05 & 0.49 & 0.60 & 0.64 & 0.77 & 0.97 \\
 & \metricfontsize NMI & 0.09 & 0.10 & 0.08 & 0.09 & 0.03 & 0.00 & 0.47 & 0.00 & \textbf{0.67} & 0.41 & 0.01 & 0.01 & 0.32 & 0.26 & 0.01 & 0.36 & 0.47 & 0.51 & 0.48 & 0.48 \\
\hline
\multirow{3}{*}{\rotatebox{\rotationangledataset}{\datasetfontsize Egypt}} & \metricfontsize Homog. & 0.05 & 0.01 & 0.02 & 0.11 & 0.00 & 0.19 & 0.92 & \textbf{1.00} & \textbf{1.00} & \textbf{1.00} & 0.00 & 0.00 & 0.17 & 0.15 & 0.64 & 0.39 & 0.91 & 0.53 & 0.00 & \textbf{1.00} \\
 & \metricfontsize WP & 0.43 & 0.40 & 0.41 & 0.47 & 0.00 & 0.54 & 0.96 & \textbf{1.00} & \textbf{1.00} & \textbf{1.00} & 0.00 & 0.00 & 0.93 & 0.93 & 0.84 & 0.77 & 0.99 & 0.89 & 0.00 & \textbf{1.00} \\
 & \metricfontsize NMI & 0.08 & 0.02 & 0.04 & 0.16 & 0.00 & 0.19 & 0.61 & 0.03 & 0.62 & 0.19 & 0.00 & 0.00 & 0.30 & 0.33 & 0.58 & 0.34 & \textbf{0.85} & 0.43 & 0.00 & 0.46 \\
\hline
\multirow{3}{*}{\rotatebox{\rotationangledataset}{\datasetfontsize Ghana}} & \metricfontsize Homog. & 0.08 & 0.09 & 0.09 & 0.20 & 0.00 & 0.16 & 0.88 & \textbf{1.00} & 0.97 & \textbf{1.00} & 0.00 & 0.00 & 0.15 & 0.15 & 0.53 & 0.21 & 0.71 & 0.29 & 0.00 & 0.98 \\
 & \metricfontsize WP & 0.14 & 0.16 & 0.16 & 0.28 & 0.00 & 0.12 & 0.84 & \textbf{1.00} & 0.94 & \textbf{1.00} & 0.00 & 0.00 & 0.25 & 0.25 & 0.44 & 0.20 & 0.69 & 0.31 & 0.00 & 0.97 \\
 & \metricfontsize NMI & 0.13 & 0.15 & 0.15 & 0.28 & 0.00 & 0.00 & 0.57 & 0.18 & 0.53 & 0.05 & 0.00 & 0.00 & 0.32 & 0.32 & 0.35 & 0.00 & 0.47 & 0.05 & 0.00 & \textbf{0.67} \\
\hline
\multirow{3}{*}{\rotatebox{\rotationangledataset}{\datasetfontsize Iran 1}} & \metricfontsize Homog. & 0.03 & 0.04 & 0.06 & 0.12 & 0.00 & 0.05 & 0.93 & \textbf{1.00} & \textbf{1.00} & \textbf{1.00} & \textbf{1.00} & 0.00 & 0.15 & 0.14 & 0.44 & 0.45 & 0.96 & 0.60 & 0.84 & \textbf{1.00} \\
 & \metricfontsize WP & 0.15 & 0.16 & 0.19 & 0.25 & 0.12 & 0.15 & 0.90 & \textbf{1.00} & \textbf{1.00} & \textbf{1.00} & \textbf{1.00} & 0.00 & 0.00 & 0.00 & 0.54 & 0.50 & 0.98 & 0.69 & 0.89 & \textbf{1.00} \\
 & \metricfontsize NMI & 0.05 & 0.07 & 0.10 & 0.18 & 0.00 & 0.00 & 0.47 & 0.05 & 0.38 & 0.04 & 0.00 & 0.00 & 0.32 & 0.32 & 0.50 & 0.21 & \textbf{0.93} & 0.46 & 0.74 & 0.23 \\
\hline
\multirow{3}{*}{\rotatebox{\rotationangledataset}{\datasetfontsize Iran 2}} & \metricfontsize Homog. & 0.00 & 0.02 & 0.05 & 0.07 & 0.00 & 0.01 & 0.87 & \textbf{1.00} & 0.97 & \textbf{1.00} & \textbf{1.00} & 0.00 & 0.14 & 0.14 & 0.44 & 0.61 & 0.91 & 0.63 & \textbf{1.00} & \textbf{1.00} \\
 & \metricfontsize WP & 0.06 & 0.08 & 0.12 & 0.14 & 0.00 & 0.06 & 0.79 & 0.00 & 0.91 & \textbf{1.00} & 0.00 & 0.00 & 0.00 & 0.00 & 0.48 & 0.60 & 0.94 & 0.63 & 0.00 & 0.99 \\
 & \metricfontsize NMI & 0.00 & 0.04 & 0.09 & 0.12 & 0.00 & 0.00 & 0.46 & 0.00 & 0.40 & 0.08 & 0.00 & 0.00 & 0.32 & 0.32 & 0.47 & 0.55 & 0.88 & 0.53 & \textbf{1.00} & 0.38 \\
\hline
\multirow{3}{*}{\rotatebox{\rotationangledataset}{\datasetfontsize Iran 3}} & \metricfontsize Homog. & 0.00 & 0.02 & 0.04 & 0.11 & 0.00 & 0.02 & 0.96 & \textbf{1.00} & 0.98 & \textbf{1.00} & \textbf{1.00} & 0.00 & 0.14 & 0.14 & 0.33 & 0.51 & 0.83 & 0.45 & \textbf{1.00} & \textbf{1.00} \\
 & \metricfontsize WP & 0.09 & 0.11 & 0.14 & 0.21 & 0.00 & 0.10 & 0.95 & 0.00 & 0.97 & \textbf{1.00} & 0.00 & 0.00 & 0.00 & 0.00 &  0.37 & 0.41 & 0.88 & 0.50 & 0.00 & \textbf{1.00} \\
 & \metricfontsize NMI & 0.00 & 0.05 & 0.08 & 0.17 & 0.00 & 0.00 & 0.62 & 0.00 & 0.55 & 0.11 & 0.00 & 0.00 & 0.32 & 0.32 & 0.28 & 0.11 & 0.63 & 0.24 & \textbf{1.00} & 0.26 \\
\hline
\multirow{3}{*}{\rotatebox{\rotationangledataset}{\datasetfontsize Iran 4}} & \metricfontsize Homog. & 0.00 & 0.00 & 0.01 & 0.02 & 0.00 & 0.06 & 0.97 & \textbf{1.00} & \textbf{1.00} & \textbf{1.00} & \textbf{1.00} & \textbf{1.00} & 0.13 & 0.13 & 0.77 & 0.76 & 0.91 & 0.77 & \textbf{1.00} & \textbf{1.00} \\
 & \metricfontsize WP & 0.25 & 0.25 & 0.25 & 0.26 & 0.00 & 0.30 & 0.98 & \textbf{1.00} & \textbf{1.00} & \textbf{1.00} & 0.00 & 0.00 & 0.00 & 0.00 & 0.86 & 0.88 & 0.95 & 0.88 & 0.00 & 0.99 \\
 & \metricfontsize NMI & 0.00 & 0.00 & 0.01 & 0.03 & 0.00 & 0.00 & 0.48 & 0.00 & 0.68 & 0.02 & 0.00 & 0.00 & 0.32 & 0.32 & 0.76 & 0.71 & 0.84 & 0.63 & \textbf{1.00} & 0.12 \\
\hline
\multirow{3}{*}{\rotatebox{\rotationangledataset}{\datasetfontsize Iran 5}} & \metricfontsize Homog. & 0.00 & 0.00 & 0.00 & 0.01 & 0.00 & 0.01 & 0.94 & \textbf{1.00} & \textbf{1.00} & \textbf{1.00} & 0.00 & 0.00 & 0.13 & 0.13 & 0.85 & 0.83 & 0.96 & 0.62 & 0.00 & 0.98 \\
 & \metricfontsize WP & 0.00 & 0.00 & 0.08 & 0.08 & 0.00 & 0.08 & 0.94 & 0.00 & \textbf{1.00} & \textbf{1.00} & 0.00 & 0.00 & 0.00 & 0.00 & 0.85 & 0.87 & 0.96 & 0.49 & 0.00 & 0.94 \\
 & \metricfontsize NMI & 0.00 & 0.00 & 0.00 & 0.01 & 0.00 & 0.00 & 0.75 & 0.00 & \textbf{0.94} & 0.52 & 0.00 & 0.00 & 0.32 & 0.32 & 0.75 & 0.70 & \textbf{0.94} & 0.00 & 0.00 & 0.24 \\
\hline
\multirow{3}{*}{\rotatebox{\rotationangledataset}{\datasetfontsize Iran 6}} & \metricfontsize Homog. & 0.04 & 0.04 & 0.07 & 0.11 & 0.00 & 0.12 & 0.85 & \textbf{1.00} & 0.93 & \textbf{1.00} & \textbf{1.00} & \textbf{1.00} & 0.13 & 0.12 & 0.25 & 0.63 & 0.59 & 0.63 & \textbf{1.00} & 0.99 \\
 & \metricfontsize WP & 0.04 & 0.06 & 0.09 & 0.13 & 0.00 & 0.03 & 0.77 & \textbf{1.00} & 0.81 & \textbf{1.00} & 0.00 & 0.00 & 0.00 & 0.00 & 0.14 & 0.56 & 0.39 & 0.49 & 0.00 & 0.99 \\
 & \metricfontsize NMI & 0.00 & 0.07 & 0.13 & 0.17 & 0.00 & 0.00 & 0.64 & 0.07 & 0.49 & 0.11 & 0.00 & 0.00 & 0.32 & 0.33 & 0.16 & 0.46 & 0.34 & 0.40 & \textbf{1.00} & 0.65 \\
\hline
\multirow{3}{*}{\rotatebox{\rotationangledataset}{\datasetfontsize Qatar}} & \metricfontsize Homog. & 0.00 & 0.00 & 0.00 & 0.05 & 0.00 & 0.03 & 0.67 & \textbf{1.00} & 0.98 & \textbf{1.00} & \textbf{1.00} & 0.00 & 0.08 & 0.08 & 0.33 & 0.20 & 0.33 & 0.30 & \textbf{1.00} & 0.87 \\
 & \metricfontsize WP & 0.00 & 0.00 & 0.00 & 0.01 & 0.00 & 0.00 & 0.21 & 0.00 & \textbf{0.91} & 0.00 & 0.00 & 0.00 & 0.01 & 0.01 & 0.23 & 0.03 & 0.04 & 0.05 & 0.00 & 0.52 \\
 & \metricfontsize NMI & 0.00 & 0.00 & 0.00 & 0.00 & 0.00 & 0.00 & 0.13 & 0.00 & 0.46 & 0.00 & 0.00 & 0.00 & 0.33 & 0.34 & 0.34 & 0.00 & 0.00 & 0.00 & \textbf{1.00} & 0.32 \\
\hline
\multirow{3}{*}{\rotatebox{\rotationangledataset}{\datasetfontsize Russia 1}} & \metricfontsize Homog. & 0.01 & 0.00 & 0.01 & 0.05 & 0.00 & 0.07 & 0.94 & \textbf{1.00} & \textbf{1.00} & \textbf{1.00} & \textbf{1.00} & \textbf{1.00} & 0.06 & 0.06 & 0.39 & 0.83 & 0.98 & 0.88 & 0.76 & 0.99 \\
 & \metricfontsize WP & 0.10 & 0.10 & 0.11 & 0.15 & 0.10 & 0.14 & 0.94 & 0.00 & \textbf{1.00} & 0.00 & 0.98 & 0.00 & 0.00 & 0.00 & 0.45 & 0.84 & 0.99 & 0.90 & 0.88 & 0.99 \\
 & \metricfontsize NMI & 0.00 & 0.01 & 0.02 & 0.07 & 0.01 & 0.00 & 0.77 & 0.00 & 0.63 & 0.00 & 0.01 & 0.00 & 0.34 & 0.35 & 0.36 & 0.71 & \textbf{0.98} & 0.81 & 0.62 & 0.69 \\
\hline
\multirow{3}{*}{\rotatebox{\rotationangledataset}{\datasetfontsize Russia 2}} & \metricfontsize Homog. & 0.00 & 0.00 & 0.00 & 0.01 & 0.00 & 0.06 & 0.94 & \textbf{1.00} & 0.99 & \textbf{1.00} & 0.00 & 0.00 & 0.05 & 0.05 & 0.48 & 0.66 & 0.89 & 0.43 & \textbf{1.00} & \textbf{1.00} \\
 & \metricfontsize WP & 0.14 & 0.15 & 0.15 & 0.16 & 0.00 & 0.19 & 0.92 & 0.00 & 0.96 & \textbf{1.00} & 0.00 & 0.00 & 0.00 & 0.00 & 0.55 & 0.70 & 0.93 & 0.47 & \textbf{1.00} & 0.97 \\
 & \metricfontsize NMI & 0.00 & 0.01 & 0.01 & 0.02 & 0.00 & 0.00 & 0.56 & 0.00 & 0.36 & 0.03 & 0.00 & 0.00 & 0.35 & 0.36 & 0.34 & 0.56 & 0.79 & 0.40 & \textbf{1.00} & 0.15 \\
\hline
\multirow{3}{*}{\rotatebox{\rotationangledataset}{\datasetfontsize Russia 3}} & \metricfontsize Homog. & 0.00 & 0.00 & 0.00 & 0.00 & 0.00 & 0.02 & 0.87 & \textbf{1.00} & \textbf{1.00} & \textbf{1.00} & 0.00 & 0.00 & 0.04 & 0.05 & \textbf{1.00} & 0.21 & 0.76 & 0.47 & 0.00 & 0.77 \\
 & \metricfontsize WP & 0.00 & 0.00 & 0.01 & 0.01 & 0.00 & 0.01 & 0.17 & 0.00 & 0.00 & 0.00 & 0.00 & 0.00 & 0.00 & 0.00 & 0.00 & 0.07 & \textbf{0.50} & 0.13 & 0.00 & 0.42 \\
 & \metricfontsize NMI & 0.00 & 0.00 & 0.00 & 0.00 & 0.00 & 0.00 & 0.00 & 0.00 & 0.00 & 0.00 & 0.00 & 0.00 & 0.36 & 0.37 & \textbf{1.00} & 0.00 & 0.55 & 0.00 & 0.00 & 0.36 \\
\hline
\multirow{3}{*}{\rotatebox{\rotationangledataset}{\datasetfontsize Russia 4}} & \metricfontsize Homog. & 0.00 & 0.07 & 0.04 & 0.05 & 0.00 & 0.08 & 0.58 & \textbf{1.00} & 0.91 & \textbf{1.00} & \textbf{1.00} & \textbf{1.00} & 0.04 & 0.04 & 0.10 & 0.23 & 0.54 & 0.35 & 0.67 & 0.88 \\
 & \metricfontsize WP & 0.00 & 0.08 & 0.00 & 0.00 & 0.00 & 0.00 & 0.03 & 0.00 & \textbf{0.58} & 0.00 & 0.00 & 0.00 & 0.00 & 0.00 & 0.00 & 0.01 & 0.18 & 0.04 & 0.40 & 0.42 \\
 & \metricfontsize NMI & 0.00 & 0.13 & 0.00 & 0.00 & 0.00 & 0.00 & 0.00 & 0.00 & 0.36 & 0.00 & 0.00 & 0.00 & \textbf{0.37} & \textbf{0.37} & 0.00 & 0.00 & 0.24 & 0.00 & 0.31 & 0.28 \\
\hline
\multirow{3}{*}{\rotatebox{\rotationangledataset}{\datasetfontsize Russia 5}} & \metricfontsize Homog. & 0.05 & 0.00 & 0.00 & 0.02 & 0.00 & 0.05 & 0.59 & \textbf{1.00} & \textbf{1.00} & \textbf{1.00} & 0.00 & 0.00 & 0.03 & 0.04 & 0.47 & 0.17 & 0.64 & 0.27 & 0.24 & 0.98 \\
 & \metricfontsize WP & 0.03 & 0.00 & 0.00 & 0.01 & 0.00 & 0.01 & 0.16 & 0.00 & \textbf{1.00} & 0.00 & 0.00 & 0.00 & 0.00 & 0.00 & 0.31 & 0.04 & 0.33 & 0.06 & 0.33 & 0.88 \\
 & \metricfontsize NMI & 0.00 & 0.00 & 0.00 & 0.00 & 0.00 & 0.00 & 0.09 & 0.00 & 0.14 & 0.00 & 0.00 & 0.00 & 0.37 & \textbf{0.38} & \textbf{0.38} & 0.03 & 0.16 & 0.04 & 0.00 & 0.37 \\
\hline
\multirow{3}{*}{\rotatebox{\rotationangledataset}{\datasetfontsize Spain}} & \metricfontsize Homog. & 0.04 & 0.05 & 0.16 & 0.19 & 0.00 & 0.35 & 0.90 & 0.99 & 0.82 & \textbf{1.00} & 0.00 & 0.00 & 0.03 & 0.03 & 0.76 & 0.65 & 0.64 & 0.34 & 0.00 & 0.99 \\
 & \metricfontsize WP & 0.16 & 0.18 & 0.29 & 0.33 & 0.00 & 0.41 & 0.94 & 0.99 & 0.85 & \textbf{1.00} & 0.00 & 0.00 & 0.32 & 0.32 & 0.81 & 0.78 & 0.69 & 0.49 & 0.00 & 0.99 \\
 & \metricfontsize NMI & 0.07 & 0.09 & 0.23 & 0.27 & 0.00 & 0.30 & \textbf{0.83} & 0.46 & 0.69 & 0.31 & 0.00 & 0.00 & 0.35 & 0.33 & 0.72 & 0.62 & 0.57 & 0.00 & 0.00 & 0.71 \\
\hline
\multirow{3}{*}{\rotatebox{\rotationangledataset}{\datasetfontsize Thailand}} & \metricfontsize Homog. & 0.00 & 0.00 & 0.00 & 0.00 & 0.00 & 0.12 & 0.96 & \textbf{1.00} & 0.96 & \textbf{1.00} & 0.00 & 0.00 & 0.04 & 0.03 & 0.61 &  0.35 & 0.85 & 0.44 & 0.00 & 0.96 \\
 & \metricfontsize WP & 0.15 & 0.00 & 0.15 & 0.15 & 0.00 & 0.26 & 0.94 & 0.00 & 0.96 & \textbf{1.00} & 0.00 & 0.00 & 0.37 & 0.11 & 0.64 & 0.47 & 0.90 & 0.65 & 0.00 & 0.95 \\
 & \metricfontsize NMI & 0.00 & 0.00 & 0.00 & 0.00 & 0.00 & 0.00 & 0.42 & 0.00 & 0.48 & 0.08 & 0.00 & 0.00 & 0.32 & 0.30 & 0.45 & 0.26 & \textbf{0.83} & 0.36 & 0.00 & 0.62 \\
\hline
\multirow{3}{*}{\rotatebox{\rotationangledataset}{\datasetfontsize Uae}} & \metricfontsize Homog. & 0.25 & 0.01 & 0.03 & 0.08 & 0.00 & 0.25 & 0.96 & \textbf{1.00} & 0.99 & \textbf{1.00} & \textbf{1.00} & \textbf{1.00} & 0.09 & 0.03 & 0.58 & 0.74 & 0.94 & 0.65 & \textbf{1.00} & \textbf{1.00} \\
 & \metricfontsize WP & 0.48 & 0.26 & 0.29 & 0.34 & 0.26 & 0.48 & 0.98 & \textbf{1.00} & \textbf{1.00} & \textbf{1.00} & \textbf{1.00} & \textbf{1.00} & 0.68 & 0.68 & 0.73 & 0.87 & 0.97 & 0.84 & \textbf{1.00} & \textbf{1.00} \\
 & \metricfontsize NMI & 0.32 & 0.01 & 0.06 & 0.13 & 0.00 & 0.24 & 0.78 & 0.29 & 0.73 & 0.14 & 0.00 & 0.00 & 0.40 & 0.43 & 0.48 & 0.63 & 0.88 & 0.53 & \textbf{1.00} & 0.63 \\
\hline
\multirow{3}{*}{\rotatebox{\rotationangledataset}{\datasetfontsize Venezuela 1}} & \metricfontsize Homog. & 0.00 & 0.00 & 0.00 & 0.00 & 0.00 & 0.22 & 0.90 & \textbf{1.00} & \textbf{1.00} & \textbf{1.00} & \textbf{1.00} & 0.00 & 0.02 & 0.02 & 0.70 & 0.37 & 0.92 & 0.15 & 0.47 & 0.99 \\
 & \metricfontsize WP & 0.08 & 0.08 & 0.08 & 0.08 & 0.00 & 0.23 & 0.91 & 0.99 & \textbf{1.00} & 0.00 & \textbf{1.00} & 0.00 & 0.00 & 0.00 & 0.67 & 0.39 & 0.89 & 0.06 & 0.95 & 0.92 \\
 & \metricfontsize NMI & 0.00 & 0.00 & 0.00 & 0.00 & 0.00 & 0.00 & 0.81 & 0.28 & \textbf{0.90} & 0.00 & 0.01 & 0.00 & 0.43 & 0.44 & 0.60 & 0.15 & 0.86 & 0.00 & 0.00 & 0.17 \\
\hline
\multirow{3}{*}{\rotatebox{\rotationangledataset}{\datasetfontsize Venezuela 2}} & \metricfontsize Homog. & 0.04 & 0.00 & 0.00 & 0.00 & 0.00 & 0.26 & 0.57 & \textbf{1.00} & 0.93 & \textbf{1.00} & \textbf{1.00} & 0.00 & 0.02 & 0.02 & 0.60 & 0.16 & 0.94 & 0.53 & \textbf{1.00} & 0.81 \\
 & \metricfontsize WP & 0.04 & 0.00 & 0.01 & 0.01 & 0.00 & 0.04 & 0.27 & \textbf{1.00} & 0.83 & \textbf{1.00} & \textbf{1.00} & 0.00 & 0.00 & 0.00 & 0.30 & 0.06 & 0.90 & 0.40 & \textbf{1.00} & 0.64 \\
 & \metricfontsize NMI & 0.06 & 0.00 & 0.00 & 0.00 & 0.00 & 0.00 & 0.00 & 0.63 & 0.62 & 0.17 & 0.09 & 0.00 & 0.44 & 0.45 & 0.14 & 0.00 & 0.84 & 0.37 & \textbf{1.00} & 0.58 \\

\bottomrule
\end{tabular}
\caption{Performance of Coordination Detection Approaches on Various Datasets.}
\label{table:results_h_wp}

\end{table*}

\end{document}